\documentclass{llncs}
\usepackage[utf8]{inputenc}

\usepackage{amsmath,amssymb}
\usepackage{graphicx}
\usepackage{xspace}
\usepackage{xcolor}
\usepackage{tabularx}
\usepackage{enumitem}
\usepackage{booktabs}
\usepackage{cite}

\usepackage[caption=false]{subfig}

\newtheorem{defn}[theorem]{Definition}

\def\leq{\leqslant}
\def\geq{\geqslant}

\DeclareMathOperator{\loglog}{loglog}

\def\topgrad_#1{\widetilde \nabla\!_{#1}}

\newcommand{\topnab}{\mathop{\widetilde \triangledown}}

\newcommand{\mc}{\mathcal}

\newcommand{\N}{\ensuremath{\mathbf{N}}\xspace}	
\newcommand{\R}{\ensuremath{\mathbf{R}}\xspace}	

\newcommand*{\etc}{etc.\@\xspace}
\newcommand*{\ie}{i.e.\@\xspace}
\newcommand*{\cf}{cf.\@\xspace}

\newcommand*{\aas}{a.a.s.\@\xspace}
\newcommand*{\whp}{w.h.p.\@\xspace}

\def\Prob{\mathbb{P}} 

\usepackage{microtype}

\title{Hyperbolicity, degeneracy, and expansion of random intersection graphs}

\author{Matthew Farrell\inst{1}, Timothy D.~Goodrich\inst{2}, Nathan Lemons\inst{3}, Felix
  Reidl\inst{2},\\ Fernando S\'anchez Villaamil\inst{4} and Blair D.~Sullivan\inst{2}}

\authorrunning{M.~Farrell, T.D.~Goodrich, N.~Lemons, F.~Reidl, F.~S\'anchez and B.D.~Sullivan} 
\institute{
Department of Mathematics, Cornell University\\
Ithaca, NY, U.S.A.\\
\email{msf235@cornell.edu}
\and
Department of Computer Science, North Carolina State University\\
Raleigh, NC, U.S.A.\\
\email{\{tdgoodri,fjreidl,blair\_sullivan\}@ncsu.edu}
\and
Theoretical Division, Los Alamos National Laboratory\\
Los Alamos, NM, U.S.A.\\
\email{nlemons@gmail.com}
\and
Theoretical Computer Science, RWTH Aachen\\
Aachen, Germany\\
\email{\{fernando.sanchez\}@cs.rwth-aachen.de}
}

\begin{document}

\maketitle

\begin{abstract}
  We establish the conditions under which several algorithmically exploitable
  structural features hold for random intersection graphs, a natural model for
  many real-world networks where edges correspond to shared attributes.
  Specifically, we fully characterize the degeneracy of random intersection
  graphs, and prove that the model asymptotically almost surely produces
  graphs with hyperbolicity at least $\log{n}$. Further, we prove that in the parametric
  regime where random intersection graphs are degenerate an even stronger
  notion of sparseness, so called \emph{bounded expansion}, holds with high
  probability.

  We supplement our theoretical findings with experimental evaluations of the
  relevant statistics.
\end{abstract}

\section{Introduction}

There has been a recent surge of interest in analyzing large graphs, stemming
from the rise in popularity (and scale) of social networks and significant
growth of relational data in science and engineering fields (e.g. gene
expressions, cybersecurity logs, and neural connectomes). One significant
challenge in the field is the lack of deep understanding of the underlying
structure of various classes of real-world networks. Here, we focus on two
structural characteristics that can be exploited algorithmically:
\emph{bounded expansion}\footnote{Not related to the notion of expander
graphs.} and \emph{hyperbolicity}.

A graph class has \emph{bounded expansion} if, for every member~$G$, one cannot
form arbitrarily dense graphs by contracting subgraphs of small radius.
Formally, the degeneracy of every minor of~$G$ is bounded by a function of the
\emph{depth} of that minor (the maximum radius of its branch sets).  Bounded
expansion offers a structural generalization of both bounded-degree and graphs
excluding a (topological) minor.  Algorithmically, this property is extremely
useful: every first-order-definable problem is decidable in linear fpt-time in
these classes~\cite{DvorakKT13}. We also consider
\emph{$\delta$-hyperbolicity}, which restricts the structure of shortest-path
distances in the graph to be tree-like. Hyperbolicity is closely tied to
treelength, but unrelated to measures of structural density such as bounded
expansion. Algorithms for graph classes of bounded hyperbolicity often exploit
computable approximate distance trees~\cite{Chepoi08} or greedy
routing~\cite{Kleinberg07}. Both of these properties present challenges for
empirical evaluation---bounded expansion is only defined with respect to graph
classes (not for single instances), and hyperbolicity is an extremal statistic
whose O($n^4$) computation is infeasible for many of today's large data sets.
As is typical in the study of network structure, we instead ask how the
properties behave with respect to randomized models which are designed to
mimic aspects of network formation and structure.

In this paper, we consider the \emph{random intersection graph model}
introduced by Karo\'nski, Scheinerman, and Singer-Cohen \cite{sin,kss} which
has recently attracted significant attention in the
literature~\cite{bloznelis-2013-assortativity,dk,jks,gjr1,ryb-inhomogenous}.
\emph{Random intersection graphs} are based on the premise that network edges
often represent underlying shared interests or attributes. The model first
creates a bipartite object-attribute graph $B = (V,A,E)$ by adding edges
uniformly at random with a fixed per-edge probability~$p(\alpha)$, then
considers the \emph{intersection graph} ~$G := (V,E')$, where $xy \in E'$ if and only if
the neighborhoods of the vertices~$x,y$ in~$B$ have a non-empty intersection.
The parameter~$\alpha$ controls both the ratio of attributes to objects and
the probability~$p$: For~$n$ objects, the number of attributes~$m$ is
proportional to~$n^\alpha$ and the probability~$p$ to~$n^{-(1+\alpha)/2}$.

Random intersection graphs are particularly attractive because they meet three
important criteria: (1) the generative process makes sense in many real-world
contexts, for example collaboration networks of
scientists~\cite{watts-1998-collective,newman-2001-scientific}; (2) they are
able to generate graphs which match key empirically established properties of
real data---namely sparsity, (tunable) clustering and
assortativity~\cite{dk,bloznelis-2013-degree,bloznelis-2013-assortativity};
and (3) they are relatively mathematically tractable due to significant amounts
of independence in the underlying edge creation process.  In this paper, we
present the following results on the structure of random intersection graphs.
\begin{enumerate}[label={(\roman{enumi})}]
\item For $\alpha\leq1$, with high probability (w.h.p.), random intersection graphs are
  somewhere dense (and thus do not have bounded expansion) and have
  unbounded degeneracy.
\item For $\alpha>1$, w.h.p.\, random intersection graphs have bounded
  expansion (and thus constant degeneracy).
\item Under reasonable restrictions on the constants in the model, random
  intersection graphs have hyperbolicity $\Omega(\log n)$ asymptotically almost
  surely.
\end{enumerate}

\noindent
In particular, the second result strengthens the original claim that the model
generates sparse graphs for $\alpha > 1$, by establishing they are in fact
\emph{structurally} sparse in a robust sense.
It is of interest to note that random intersection graphs only exhibit
tunable clustering when $\alpha = 1$~\cite{dk}, when our results
indicate they are not structurally sparse (in any reasonable
sense)\footnote{This is not tautological---a result
  in~\cite{BndExpNetw14} shows that constant clustering and bounded
  expansion are not orthogonal.}.
Further, we note that the third result is negative---our bound implies a
$\log{n}$ lower bound on the treelength~\cite{Chepoi08}.

\section{Preliminaries}\label{sec:preliminaries}

We start with a few necessary definitions and lemmas, covering each of
the key ideas in the paper (random intersection graphs, degeneracy,
expansion, and hyperbolicity). We use standard notation: $G=(V,E)$
denotes a finite, simple graph on the vertices $V$ with edge set $E$.
We alternatively write~$V(G)$ and~$E(G)$ to denote the edge and vertex
set, respectively.
For a graph $G$ and a vertex $x\in V(G)$, $N_G(x)$ denotes the set
of neighbors of $x$ in $G$. A subgraph $H$ of $G$
is {\em induced} if for every pair of vertices $u,v \in
V(H)$, the edge $(u,v)$ exists in $H$ if and only if it exists in~$G$. Paths and
cycles consisting of $k$ edges are said to have {\it length} $k$ and
are denoted $P_k$ and $C_k$ respectively.  For vertices $x$ and $y$ in
a graph let $P[x,y]$ denote a shortest path from $x$ to $y$.

We use
the terms \emph{asymptotically almost surely} (\aas) and \emph{with high
probability} (\whp) according to the following conventions:
For each integer $n$, let $\mathcal{G}_n$ define a
distribution on graphs with $n$ vertices (for example, coming from a random
graph model). We say the event $E_n$ defined on $\mathcal{G}_n$ holds
\emph{asymptotically almost surely} (\aas) if
$\lim_{n\rightarrow\infty}\mathbb{P}[E_n]= 1$.
We say an event occurs
\emph{with high probability} (\whp) if for any $c \geq 1$ the event occurs with
probability at least $1 - f(c)/n^c$ for $n$ greater than some constant,
where $f$ is some function only depending on $c$. As a shorthand, we will
simply say that $\mathcal G_n$ \emph{has some property} \aas (or \whp).

\subsection{Random Intersection Graphs}\label{sec:rig}
A wide variety of random intersection graph models have been defined in the
literature. In this paper, we restrict our attention to the most well-studied
of these models, $G(n,m,p)$, which is defined as follows:

\begin{defn}[Random Intersection Graph Model]
  Fix positive constants $\alpha,\beta$ and $\gamma$.  Let $B$ be a random
  bipartite graph on parts of size $n$ and $m=\beta n^{\alpha}$ with each edge
  present independently with probability $p=\gamma n^{-(1+\alpha)/2}$.  Let $V$
  (the vertices) denote the part of size $n$ and $A$ (the attributes) the part of
  size $m$.
  The associated  \emph{random intersection graph $G = G(n,m,p)$} is
  defined on the vertices $V$: two vertices are connected in $G$ if they share (are
  both adjacent to in $B$) at least one attribute in $A$.
\end{defn}

\noindent
We note that $G(n,m,p)$ defines a distribution~$\mathcal{G}_n$ on graphs with
$n$ vertices. The notation $G = G(n,m,p)$ denotes a graph $G$ that is randomly
sampled from the distribution $\mathcal{G}_n$. Throughout the manuscript,
given a random intersection graph $G(n,m,p)$, we will refer to $B$ as the
associated bipartite graph on $n$ vertices and $m$ attributes from which $G$ is
formed.

In order to work with graph classes formed by the random intersection graph
model, we will need a technical result that bounds the number of attributes in
the neighborhood of a subset of vertices around its expected value. These lemmas
and their proofs are in Section~\ref{sec:rig-lemmas}.

\subsection{Degeneracy \& Expansion}

\noindent
Although it is widely accepted that complex networks tend to be sparse (in
terms of edge density), this property usually is not sufficient to improve
algorithmic tractability: many NP-hard problems on graphs, for instance,
remain NP-hard when restricted to graphs with bounded average degree.
In contrast, graph classes that
are \emph{structurally} sparse (bounded treewidth, planar, \etc) often
admit more efficient algorithms---in particular when viewed through the
lens of parameterized complexity. Consequently, we are interested whether
random graph models and, by extension, real-world networks exhibit any
form of structural sparseness that might be exploitable algorithmically.

As a first step, we would like that a graph is not only sparse on average, but that
this property extends to all of its subgraphs. This requirement motivates a very general
class of structurally sparse graphs---that of bounded \emph{degeneracy}.

\begin{defn}[$k$-core]
  The {\em $k$-core} of $G$ is the maximum induced
  subgraph of $G$ in which all vertices have degree at least $k$.
  The {\em degeneracy} of $G$ is the maximum $k$ so that the $k$-core is nonempty
  (equivalently, the least positive integer $k$ such that every induced subgraph
  of $G$ contains a vertex with at most $k$ neighbors).
\end{defn}

\noindent
It is easy to see that the degeneracy is lower-bounded by the size of the
largest clique. Thus, the degeneracy of intersection graphs is bounded below
by the maximum attribute degree in the associated bipartite graph since each
attribute contributes a complete subgraph of size equal to its degree to the
intersection graph.  For certain parameter values, this lower bound will,
\whp, give the correct order of magnitude of the degeneracy of the graph.

\begin{figure}[t]
  \centering\includegraphics[scale=.8]{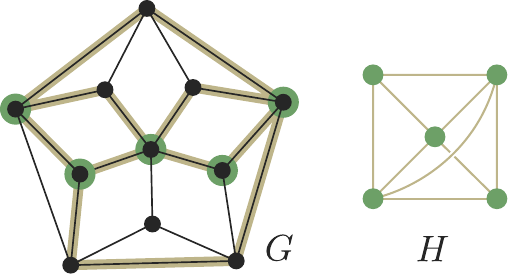}
  \caption{\label{fig:shallowminor}%
    The graph~$H$ on the right is a $1$-shallow topological minor of~$G$, as
    witnessed by the $\leq 2$-subdivision highlighted inside~$G$. Further, $H$
    is the densest among all $1$-shallow topological minor of~$G$:
    hence~$\topgrad_1(G) = |E(H)|/|V(H)| = 9/5$. }
\end{figure}

Some classes of graphs with bounded degeneracy have stronger structural
properties---here we focus on the so-called graphs of \emph{bounded
expansion}~\cite{NOdM08}. In the context of networks, bounded expansion
captures the idea that networks decompose into small dense structures (e.g.
communities) connected by a sparse global structure. More formally, we
characterize bounded-expansion classes using special graph minors and an
associated density measure `grad' (\cf Figure~\ref{fig:shallowminor}).

\begin{defn}[Shallow topological minor, nails, subdivision vertices]\label{def:shallowtopminor+}
  A graph $M$ is an \emph{$r$-shallow topological minor of~$G$} if a
  $(\leq 2r)$-subdivision of $M$ is isomorphic to a subgraph $G'$ of~$G$.
  We call $G'$ a \emph{model of $M$ in $G$}. For simplicity, we assume by
  default that $V(M) \subseteq V(G')$ such that the isomorphism between $M$
  and $G'$ is the identity when restricted to $V(M)$. The vertices $V(M)$ are
  called \emph{nails} and the vertices $V(G') \setminus V(M)$
  \emph{subdivision vertices}.
  The set of all $r$-shallow topological minors of a graph $G$ is denoted by
  $G \topnab r$.
\end{defn}

\begin{defn}[Topological grad]
  For a graph $G$ and integer $r \geq 0$, the
  \emph{topological greatest reduced average density (grad) at depth $r$} is defined as
  $
    \topgrad_r(G) = \max_{H \in G \topnab r} |E(H)|/|V(H)|.
  $
  For a graph class $\mc G$, define $\topgrad_r(\mc G) = \sup_{G \in \mc G} \topgrad_r(G)$.
\end{defn}

\begin{defn}[Bounded expansion]
  A graph class $\mc G$ has \emph{bounded expansion} if there exists a function
  $f$ such that for all $r$, we have $\topgrad_r(\mc G) < f(r)$.
\end{defn}

\noindent
When introduced, bounded expansion was originally defined using an equivalent
characterization based on the notion of \emph{shallow minors}
(\cf~\cite{NOdM08}): $H$ is a $r$-shallow minor of $G$ if $H$ can be obtained
from $G$ by contracting disjoint subgraphs of radius at most $r$. In the
context of our paper, however, the topological shallow minor variant proves
more useful, and we restrict our attention to this setting. Let us point out
that bounded expansion implies bounded degeneracy, with $2f(0)$ being an upper
bound on the degeneracy of the graphs.

Finally, \emph{nowhere dense} is a generalization of bounded expansion in
which we measure the \emph{clique number} instead of the edge density of
shallow minors. Let $\omega(G)$ denote the size of the largest complete
subgraph of a graph~$G$ and let $\omega(\mathcal G) = \sup_{G \in \mathcal G}
\omega(G)$ be the natural extension to graph classes~$\mathcal G$.

\begin{defn}[Nowhere dense~\cite{NOdM10,NOdM11}]
  A graph class $\mathcal G$ is \emph{nowhere dense} if there exists a
  function $f$ such that for all $r \in \N$ it holds that
  $\omega(\mathcal G \topnab r) < f(r)$.
\end{defn}

\noindent
See~\cite{NOdM12} for many equivalent notions.  A graph class is
\emph{somewhere dense} precisely when it is not nowhere dense. While in
general a graph class with unbounded degeneracy is not necessarily somewhere
dense, the negative proofs presented here show that members of the graph class
contain \whp~large cliques. This simultaneously implies unbounded degeneracy
and that the class is somewhere dense (as a clique is a 0-subdivision of
itself). Consequently, we prove a clear dichotomy: random intersection graphs
are either structurally sparse or somewhere dense.

\subsection{Gromov's Hyperbolicity}\label{sec:prelim:hyper}

The concept of $\delta$-hyperbolicity was introduced by Gromov in the context
of geometric group theory~\cite{Gromov87}. It captures how ``tree-like'' a
graph is in terms of its metric structure, and has received attention in the
analysis of real-world networks. We refer the reader
to~\cite{chen2013hyperbolicity,Jonckheere08,Kleinberg07,Narayan_The-large-scale_2011},
and references therein, for details on the motivating network applications.
There has been a recent surge of interest in studying the hyperbolicity of various
classes of random networks including small world
networks~\cite{chen2013hyperbolicity,shang2012lack}, Erd\H os-R\'enyi
random graphs~\cite{narayan2012lack}, and random graphs with expected
degrees~\cite{shang2013non}.

There are several ways of characterizing $\delta$-hyperbolic metric spaces,
all of which are equivalent up to constant
factors~\cite{Bridson09,Chepoi08,Gromov87}.  Since graphs are naturally
geodesic metric spaces when distance is defined using shortest paths, we will
use the definition based on $\delta$-slim triangles (originally attributed to
Rips~\cite{Bridson09,Gromov87}).

\begin{defn}[$\delta$-hyperbolicity]
  A graph $G = (V,E)$ is \emph{$\delta$-hyperbolic} if for all $x,y,z \in V$,
  for every choice of geodesic (shortest) paths between them--- denoted
  $P[x,y], P[x,z], P[y,z]$---we have
  $
    \forall v\in P[x,y],\;\exists w\in P[x,z]\cup P[z,y]:\; d_G(v,w)\leq\delta,
  $
  where $d_G(v,w)$ is shortest-path distance of $v$ to $w$ in $G$.
\end{defn}
That is, if $G$ is $\delta$-hyperbolic, then for each triple of vertices
$x,y,z$, and every choice of three shortest paths connecting them pairwise,
each point on the shortest path from $x$ to $y$ must be within distance
$\delta$ of  a point on one of the other paths. The \emph{hyperbolicity} of a
graph $G$ is the minimum $\delta \geq 0$ so that $G$ is $\delta$-hyperbolic.
Note that a trivial upper bound on the hyperbolicity is half the diameter
(this is true for any graph).

In this paper we give lower bounds for the hyperbolicity of the graphs in
$G(n,m,p)$.  We believe these bounds are asymptotically the correct order of
magnitude (e.g. also upper bounds). This would require that the the diameter
of connected components is also logarithmic in $n$, which has been shown for a
similar model \cite{Rybarczyk_Diameter_2011}.

\subsection{Concentration Results for Neighborhood Unions\label{sec:rig-lemmas}}

This section states and proves results showing that in random intersection
graphs the number of attributes in the combined neighborhood of a subset of
vertices is tightly concentrated around its expected value when $\alpha > 1$, and
has a tight lower bound when $\alpha = 1$ and the subset under consideration
is large enough.

\begin{lemma}\label{lem:isolates}
  Let $\alpha> 1$ and fix $\epsilon>0$. Then if $G(n,m,p)$ is a random
  intersection graph on vertex set $V$ and $S\subset V$, \whp
  $
     (1-\epsilon) |S|mp \leq |N_B(S)| \leq (1+\epsilon) |S|mp.
  $
\end{lemma}
\begin{proof}
  Let $\epsilon>0$.  Let $v$ be a vertex in $G$ and let $A(v)$ denote the
  number of attributes adjacent to $v$ in the associated bipartite graph $B$.
  Since each attribute is adjacent to $v$ independently with probability $p$,
  $A(v)$ has a binomial distribution and by Bernstein's inequality:
  \[
    \mathbb{P}[A(v)>(1+\epsilon)mp] \leq
    \exp - \Big(\frac{\epsilon^2mp}{2(1+\epsilon /3)}\Big)
    = o\Big(\frac{1}{n^c}\Big)
  \]
  for any fixed $c.$ By the union bound, it follows that with high
  probability $|N_B(S)|\leq (1+\epsilon) |S|mp$.

  For each attribute $a$ in $B$, let $\mathcal{I}_a$ be the indicator
  random variable equal to $1$ when $a$ has at least one neighbor in
  $S$. Since these variables are independently and identically distributed, we
  again use Bernstein's inequality:
  \begin{align*}
    \mathbb{P}\big[\textstyle\sum_a
      \mathcal{I}_a<(1-\epsilon)m\mathbb{P}[\mathcal{I}_a=1]\big]
    &\leq \exp -\Big(\frac{\epsilon^2 m\mathbb{P}[\mathcal{I}_a]}{2}\Big)\\
    &\leq \exp -\Big(\frac{\epsilon^2m(1-\epsilon)|S|p}{2}\Big)\\
    &\leq o\Big(\frac{1}{n^c}\Big)
  \end{align*}
  for any fixed $c$.  The penultimate inequality follows from
  $$
  \mathbb{P}[\mathcal{I}_a=1] = 1-(1-p)^{|S|}\geq |S|p(1-o(1)).
  $$
\qed\end{proof}

\begin{lemma}\label{lemma:neighbors}
  Let $\alpha= 1$ and fix $\eta,\epsilon>0$.
  If $G(n,m,p)$ is a random intersection graph on vertex set $V$ and $S\subset
  V$ a subset of size at least $\eta n$, then \whp it holds that
  $
     |N_B(S)| \geq (1-\epsilon) |S|mp.
  $
\end{lemma}
\begin{proof}
  Again, for each attribute $a$ in $B$, let $\mathcal{I}_a$ be the
  indicator random variable equal to $1$ when $a$ has at least one
  neighbor in $S$.
  \begin{align}
    \mathbb{P}\big[\textstyle\sum_a
      \mathcal{I}_a<(1-\epsilon)m\mathbb{P}[\mathcal{I}_a=1]\big]
    &\leq \exp -\Big(\frac{\epsilon^2 m\mathbb{P}[\mathcal{I}_a]}{2}\Big)\\
    &\leq \exp -\Big(\frac{\epsilon^2m\left(1-e^{p|S|}\right)}{2}\Big)\\
    &\leq o\Big(\frac{1}{n^c}\Big)
  \end{align}
  for any fixed $c$.  Again we use the fact that $
  \mathbb{P}[\mathcal{I}_a=1] = 1-(1-p)^{|S|}\geq 1-e^{p|S|}.  $
\qed\end{proof}

\section{Structural sparsity of random intersection graphs}\label{sec:sparsity}

In this section we will characterize a clear break in the sparsity of graphs
generated by $G(n,m,p)$, depending on whether $\alpha$ is strictly greater
than one. In each case, we analyze  (probabilistically) the degeneracy and
expansion of the generated class.\looseness-1

\begin{theorem}\label{thm:degen_main}
  Fix constants $\alpha,\beta$ and
  $\gamma$. Let $m=\beta n^{\alpha}$ and $p=\gamma
  n^{-(1+\alpha)/2}$.  Let $G = G(n,m,p)$.  Then the following hold \whp
\begin{enumerate}[label={(\roman{enumi})}]
\item If $\alpha < 1$, $G(n,m,p)$ is somewhere dense and $G$ has
  degeneracy $\Omega(\gamma n^{(1-\alpha)/2})$.
\item If $\alpha = 1$, $G(n,m,p)$ is somewhere dense and $G$ has
  degeneracy $\Omega(\frac{\log n}{\log \log n})$.
\item If $\alpha > 1$, $G(n,m,p)$ has bounded expansion and thus $G$ has degeneracy $O(1)$.
\end{enumerate}
\end{theorem}

We prove each of the three cases of Theorem~\ref{thm:degen_main} separately.

\subsection{Proof of Theorem~\ref{thm:degen_main} when $\alpha \leq
  1$\label{sec:proof-alpha-small}}

When $\alpha \leq 1$, we prove that \whp the random intersection graph model
generates graph classes with unbounded degeneracy by establishing the
existence of a high-degree attribute in the associated bipartite graph (thus
lower-bounding the clique number). The proof is divided into two lemmas, one
for $\alpha < 1$ and one for $\alpha = 1$, in which we prove different lower
bounds.

\begin{lemma}\label{lemma:alpha-smaller-one-undegenerate}
Fix constants $\alpha < 1,\beta$ and $\gamma$. If $m = \beta n^\alpha$ and $p
= \gamma n^{-(1+\alpha)/2}$, then w.h.p. $G = G(n,m,p)$ has degeneracy
$\Omega(\gamma n^{(1-\alpha)/2}$).
\end{lemma}

\begin{proof}
  Let $G = G(n,m,p)$ and $B = (V, A, E)$ be the bipartite graph
  associated with $G$. Define the random variable $X_i$ to be the
  number of vertices in $V$ connected to a particular attribute
  $a_i$. Then $X_i\sim Binom(n,p)$ and $\Prob[X_i < np - 1] \leq 1/2$,
  since the median of $X_i$ lies between $\lfloor np \rfloor$ and
  $\lceil np \rceil$.  Let $\mathcal{S}$ be the event that $|X_i| < np
  - 1$ for all $i \in [1,m]$.  Since the number of vertices attached
  to each attribute is independent,
  $$
  \Prob[\mathcal{S}] = \prod_{i=1}^{m} \left(1- \Prob[X_i \geq np -
    1]\right) \leq \left[1-(1-1/2)\right]^{m}=2^{-m}.
  $$
  Now, it follows that $\lim_{n\rightarrow \infty} \Prob[\mathcal{S}]= 0$, and
  \whp~the graph $G$ contains a clique of size $np - 1 = \gamma
  n^{(1-\alpha)/2} - 1$, and thus has degeneracy at least $np-1$.
\qed\end{proof}

\begin{corollary}\label{corr:smaller-one-sd}
  Fix constants $\alpha < 1,\beta$ and $\gamma$. If $m = \beta
  n^\alpha$ and $p = \gamma n^{-(1+\alpha)/2}$, then w.h.p. $G(n,m,p)$
  is somewhere dense.
\end{corollary}
\begin{proof}
  The proof of Lemma~\ref{lemma:alpha-smaller-one-undegenerate} shows
  that w.h.p.~a clique of size $\gamma n^{(1-\alpha)/2}$ exists
  already as a subgraph (\ie a $0$-subdivision) in every $G \in
  G(n,m,p)$.
\qed\end{proof}

The following lemma addresses the case when the attributes grow at the same
rate as the number of vertices. We note that Bloznelis and Kurauskas
independently proved a similar result (using a slightly different RIG model)
in \cite{bloznelis-13-cliques}; we include a slightly more direct proof here for
completeness.

\begin{lemma}\label{lem:degen_equals}
  Fix constants $\alpha = 1,\beta$ and $\gamma$. Then a random graph $G = G(n,m,p)$ has
  degeneracy $\Omega(\frac{\log n}{\log \log n})$ \whp
\end{lemma}
\begin{proof}
  Let $c$ be any constant greater than one. We will show that for
  every $k \leq \frac{\log n}{\log\log n}$, a random graph $G \in
  G(n,m,p)$ contains a clique of size~$k$ with probability at least
  $\Omega(1 - n^{-c})$.  Fix an attribute $a$. The probability that
  $a$ has degree at least $k$ in the bipartite graph is at least the
  probability that it is \emph{exactly} $k$, hence
  \begin{align*}
    {n \choose k} p^k (1-p)^{n-k} \geq {n \choose k} p^k (1-p)^n
    \geq \frac{\gamma^k}{e^\gamma k^k}.
  \end{align*}
  We will show that this converges fast enough for $\gamma < 1$; the
  case for $\gamma \geq 1$ works analogously. Therefore the
  probability that none of the $m = \beta n $ attributes has degree
  at least $k$ is at most
  \begin{align*}
    \left( 1-\frac{\gamma^k}{e^\gamma k^k} \right)^{\beta n} \leq
    e^{-\beta n\left(\frac{\gamma^k}{e^\gamma k^k} \right)}.
  \end{align*}
  We prove that this probability is smaller than $n^{-c}$ by showing
  that
  \begin{align}\label{align:probability:large:k}
    \frac{\beta}{e^\gamma} \frac{n \gamma^k}{k^k} \geq c \cdot \log
    n,
  \end{align}
  when $k = \frac{\log n}{\log\log n}$.  Let $c' = c
  e^\gamma/\beta$. Then to show Inequality
  \ref{align:probability:large:k} holds, it is enough to show
  \begin{align*}
    n \frac{(\gamma \log \log n)^{\log n/\log\log n}}{(\log n)^{\log
        n/\log\log n}} = (\gamma \log \log n)^\frac{\log n}{\log\log
      n} \geq c' \cdot \log n.
  \end{align*}
  Comparing the functions $e^{x/\log x}$ and $c' x$, we see that for
  large enough positive~$x$,
  \begin{equation*}
    x > \log c' \log x + \log^2 x
  \end{equation*}
  and equivalently
  \begin{equation*}
    e^{x / \log x} > c' \cdot x.
  \end{equation*}
  Therefore for large enough $n$, $e^{\log n / \log \log n} > c'
  \cdot \log n$, and in particular, for $n > e^{e^{e/\gamma}}$,
  $$(\gamma \log \log n)^\frac{\log n}{\log\log n} \geq c' \cdot \log n,$$ as
  previously claimed. This shows the probability that no attribute has degree
  at least~$\log n / \log\log n$ is at most $O(n^{-c})$, and the claim
  follows.
\qed\end{proof}

\begin{corollary}\label{corr:equal-one-sd}
  Fix constants $\alpha = 1,\beta$ and $\gamma$. If $m = \beta
  n^\alpha$ and $p = \gamma n^{-(1+\alpha)/2}$, then \whp~$G(n,m,p)$
  is somewhere dense.
\end{corollary}
\begin{proof}
  Lemma~\ref{lem:degen_equals} is proven by showing that \whp~it holds that a
  clique of size $\Omega\left(\log n/(\log\log n)\right)$ exists as a subgraph
  (\ie a $0$-subdivision) in every graph in $G(n,m,p)$.
\qed\end{proof}

\subsection{Proof of Theorem~\ref{thm:degen_main} when $\alpha >
  1$\label{sec:proof-alpha-big}}

In this section, we focus on the case when $\alpha > 1$. This is the
parameter range in which the model generates sparse graphs.

Before beginning, we note that if $G(n,m,p)$ has bounded expansion
\whp, then for any $p'\leq p$ and $m'\leq m$ it follows that \whp
$G(n,m',p')$ also has bounded expansion by a simple coupling argument.
Thus we can assume without loss of generality that both $\gamma$ and
$\beta$ are greater than one. For the remainder of this section, we
fix the parameters $\gamma, \beta, \alpha > 1$, the resulting number
of attributes $m = \beta n^\alpha$ and the per-edge probability $p =
\gamma n^{-(1+\alpha)/2}$.

\subsubsection{Bounded Attribute-Degrees}
As mentioned before, for a random intersection graph to be degenerate,
the attributes of the associated bipartite graph must have bounded
degree. We prove that \whp, this necessary condition is satisfied.

\begin{lemma}\label{lemma:attr-degree-bound}
  Let $c \geq 1$ be a constant such that $2\frac{\alpha+c}{\alpha-1} >
  \beta\gamma e$. Then the probability that there exists an attribute in the
  bipartite graph associated with $G(n,m,p)$ of degree higher than
  $2\frac{\alpha+c}{\alpha-1}$ is $O(n^{-c})$.
\end{lemma}
\begin{proof}
  Taking the union bound, the probability that some attribute
  has degree larger than $d$ is upper bounded by
  \begin{align*}
    m {n \choose d} p^d & \leq \frac{\beta e^d \gamma^d}{d^d}
    \cdot \frac{n^{\alpha+d}}{n^{\frac{a+1}{2}d}},
  \end{align*}
  where the first fraction is bounded by a constant as
  soon as $d > e \beta \gamma$. Then we achieve an upper bound of
  $O(n^{-c})$ as soon as $\frac{\alpha+1}{2}d - d -\alpha > c$, or equivalently,
  $d > 2 \frac{\alpha+c}{\alpha-1}$, proving the claim. \qed
\end{proof}
This result allows us to assume for the remainder of the proof that the maximum
attribute degree is bounded.

\subsubsection{Alternative Characterization of Bounded Expansion}

We now state a characterization of bounded expansion which is often helpful in
establishing the property for classes formed by random graph models.

\begin{proposition}[\!{\rm \cite{NOdMW12,NOdM12}}]\label{prop:BoundedExpChar}
  A class~$\cal C$ of graphs has bounded expansion if and only if
  there exists real-valued functions $f_1, f_2, f_3, f_4 \colon \R
  \rightarrow \R^{+}$ such that the following two conditions hold:
\begin{enumerate}
\item[(i)] For all positive $\epsilon$ and for all graphs $G \in \cal C$ with $|V(G)| >
  f_1(\epsilon)$, it holds that
  $\frac{1}{|V(G)|} \cdot |\{v \in V(G)
  \colon \deg(v) \geq f_2(\epsilon)\}| \leq \epsilon.$
\item[(ii)] For all $r \in \N$ and for all $H \subseteq G \in \cal C$ with
  $\topgrad_r(H) > f_3(r)$, it follows that
  \[|V(H)| \geq f_4(r) \cdot |V(G)|.\]
\end{enumerate}
\end{proposition}

\noindent Intuitively, this result states that any class of graphs with bounded
expansion is characterized by two properties:
\begin{enumerate}
\item[(i)] All sufficiently large members of the class have a small
  fraction of vertices of large degree.
\item[(ii)] All subgraphs of $G \in \cal C$ whose shallow topological
  minors are sufficiently dense must necessarily span a large fraction
  of the vertices of~$G$.
\end{enumerate}

\subsubsection{Stable $r$-Subdivisions}

In order to disprove the existence of an $r$-shallow
topological minor of a certain density $\delta$, we introduce a stronger
topological structure.

\begin{defn}[Stable $r$-subdivision]
  Given graphs $G,H$ we say that \emph{$G$~contains~$H$ as a stable
    $r$-subdivision} if $G$ contains $H$ as a
  $\frac{r}{2}$-shallow topological minor with model $G'$ such that every
  path in $G'$ corresponding to an edge in $H$ has exactly length
  $r+1$ and is an induced path in $G$.
\end{defn}

\noindent A stable $r$-subdivision is by definition a shallow topological minor,
thus the existence of an $r$-subdivision of density $\delta$ implies
that $\topgrad_{\frac{r}{2}}(G) \geq \delta$. We prove that the densities
are also related in the other direction.

\begin{lemma}\label{lemma:stable-subdivisions}
  A graph~$G$ with $\topgrad_{\frac{r}{2}}(G) \geq \delta$ contains a
  stable $i$-subdivision of density at least $\delta / (r+1)$ for some
  $i \in \{0,\dots,r\}$.
\end{lemma}
\begin{proof}
  Consider a $\frac{r}{2}$-shallow topological minor~$H$ of~$G$ with density
  at least $\delta$.  Let $H' \subseteq G$ be the model of~$H$ and let
  $\lambda \colon V(H') \to V(H) \cup E(H)$ be a mapping that maps nails of
  the model to vertices of the minor and subdivision vertices of the model to
  their respective edge in the model.  Consider the preimage $\lambda^{-1}$.
  As a slight abuse of notation, we can consider $\lambda^{-1}$ as a map to
  (possibly empty) paths of $H$: indeed, we can assume that every edge of $H$
  is mapped by $\lambda^{-1}$ to an \emph{induced} path in $H'$.  If $H'$ uses
  any non-induced paths, we can replace each such path by a (shorter)
  induced path and obtain a (different) model of $H$ with the desired
  property.

  We partition the edges of~$H$ by the length of their respective paths in the model: define
  $E_\ell = \{ e \in H \mid |\lambda^{-1}(e)| = \ell \}$ for $0 \leq \ell \leq r+1$. Since
  $|E(H)| = \bigcup_{0 \leq \ell \leq r+1} |E_\ell| \geq \delta |V(H)|$, there exists
  at least one set $E_\ell$ such that $|E_\ell| \geq \delta |V(H)| / (r+1)$.
  Then the subgraph $(V(H), E_\ell)$ is a stable $\ell$-subdivision of~$G$. \qed
\end{proof}

\noindent To show that a graph has no $r$-shallow minor of density $\delta$, it now
suffices to prove that no stable $i$-subdivision of density $\delta/(2r+1)$
exists for any $i
\in \{0,\dots,2r\}$. We note that the other direction would not work,
since the existence of a stable $i$-subdivision for some $i \in
\{0,\dots,2r\}$ of density $\delta/(2r+1)$ does not imply the
existence of an $r$-shallow topological minor of density $\delta$.

We now establish the probability of having this structure in the
random intersection graph model, noting that the following structural
result is surprisingly useful, and appears to have promising
applications beyond this work. We will argue that a dense subdivision
in $G$ implies the existence of a dense subgraph in the associated
bipartite graph. We show this claim by considering the existence of a stable
$r$-subdivision where all paths are induced, which is generated by a
minimal number of attributes. Notice that if a model of some graph $H$
exists, then so does a model with these properties. This fact allows us to only
consider attributes with minimum degree two, since every edge in the
path is generated by a different attribute. This assumption is key in proving the
following theorem.

\begin{theorem}\label{theorem:subdivision-prob}
   Let $c \geq 1$ be a constant and let
  $\phi = (6 e g\beta\gamma r\delta)^{5r\delta 2/(\alpha-1)}$. The probability that
  $G(n,m,p)$ contains a stable $r$-subdivision with~$k$ nails for
  $r \geq 1$ and of density $\delta > 1$ is at most
  \begin{align*}
    r\delta k \cdot \left(\frac{\phi}{n}\right)^{\frac{\alpha-1}{2}k}
  \end{align*}
\end{theorem}
\begin{proof}
  Let us first bound the probability that the bipartite graph associated with
  $G=G(n,m,p)$ contains a dense subgraph. We will then argue that a dense
  subdivision in $G$ implies the existence of such a dense bipartite subgraph.

  Let $\Prob_{\text{dense}}(\kappa,\nu,\lambda)$ be the probability
  that there exists sets $V' \subseteq V$, $A' \subseteq A$, of size
  $\kappa$ and $\nu$ respectively, such that there exist at least
  $\lambda$ edges between vertices of $V'$ and $A'$. It is easy to see that
  this probability is bounded by
  \begin{align}\label{align:subdivision-prob}
  \Prob_{\text{dense}}(\kappa,\nu,\lambda)
  &\leq \binom{n}{\kappa} \binom{m}{\nu} \sum_{d_1,\dots,d_\nu} \prod_{i=1}^{\nu} \binom{\kappa}{d_i} p^{d_i},
  \end{align}
  where $d_1,\dots,d_\nu$ represent all possible choices of the
  degrees of $\nu$ attributes such that $\sum_{i=1}^\nu {d_i} =
  \lambda$. By Lemma~\ref{lemma:attr-degree-bound}, \whp~$d_i \leq g$ and thus
  \whp there are at most $g^{\nu}$ terms in the sum of (\ref{align:subdivision-prob}).
  Using this together with Stirling's approximation
  allows us to simplify the bound as follows:
  \begin{equation}\label{eqn:pdense}
    \Prob_{\text{dense}}(\kappa,\nu,\lambda)
    \leq
    \left(\frac{n e}{\kappa}\right)^\kappa \left(\frac{g m e}{\nu}\right)^\nu
    \left(\kappa e p\right)^\lambda
    = \frac{e^{\kappa + \nu + \lambda} g^\nu \beta^\nu \gamma^\lambda}{\nu^\nu}
       \frac{\kappa^\lambda n^{\alpha \nu + \kappa}}{\kappa^\kappa n^{\frac{\alpha+1}{2}\lambda}}.
  \end{equation}
  Consider a stable $r$-subdivision $H$ in $G$ with $k$ nails and density
  $\delta$. The model of $H$ uses exactly $k + r \delta k$ vertices of $G$.
  Let $A_H$ be a minimal set of attributes that generates the edges of the
  model of $H$ in $G$. There is at least one edge between every nail and an
  attribute in $A_H$. Furthermore, since the paths connecting the nails in the
  model are induced, every subdivision vertex has at least two edges to the
  attributes $A_H$. We conclude that there exists a bipartite subgraph with
  $\kappa = k + r \delta k$ and $\lambda = 2 r \delta k + k$. Since $A_H$ is
  minimal, every attribute of $A_H$ generates at least one edge in the model
  of $H$ and therefore $|A_H| \leq (r+1)\delta k$. Let $\delta_1 = (r \delta
  + 1)$ and $\delta_2 = (2r \delta + 1)$. By the bound in Equation \ref{eqn:pdense}, the
  probability of such a structure is at most
  \begin{align*}
    &\sum_{\nu=r\delta k/g}^{r\delta k} \Prob_{\text{dense}}(\delta_1 k,\nu,\delta_2 k)\\
    \leq~&
    \sum_{\nu=r\delta k/g}^{r\delta k}
    \frac{e^{{\delta_1 k} + \nu + {\delta_2 k}} g^{\nu} \beta^{\nu} \gamma^{{\delta_2 k}}}{\nu^{\nu} {(\delta_2 k)}^{{\delta_2 k}}}
    \frac{{(\delta_1 k)}^{{\delta_2 k}} n^{\alpha \nu + \delta_1 k}}{{(\delta_1 k)}^{{\delta_1 k}} n^{\frac{\alpha+1}{2}{\delta_2 k}}}
  \end{align*}
Let $\psi$ be the exponent of $1/n$ in a term of this sum. Then we have
  \begin{align*}
     \psi &= \left(\left(\frac{\alpha+1}{2}\right){\delta_2 k} - (\alpha \nu + \delta_1 k)\right) \\
     &= \left(\left(\frac{\alpha+1}{2}\right){(2r \delta + 1) k} - (\alpha \nu + (r \delta + 1) k)\right).
     \end{align*}
Simplifying, we see that
    \begin{equation*}
   \psi = (\alpha + 1)r\delta k + \frac{\alpha + 2}{2}k - \alpha\nu - (r\delta + 1)k
    =\frac{\alpha - 1}{2} k + \alpha(r\delta k - \nu).
  \end{equation*}
  Thus we can rewrite the previous inequality as
  \begin{align*}
    &\sum_{\nu=r\delta k/g}^{r\delta k} \Prob_{\text{dense}}(\delta_1 k,\nu,\delta_2 k)\\
    \leq{} &
    \sum_{\nu=r\delta k/g}^{r\delta k}
    \frac{e^{{\delta_1 k} + \nu + {\delta_2 k}} g^{\nu} \beta^{\nu} \gamma^{{\delta_2 k}} {\delta_1}^{{\delta_2 k}}}{{\delta_1}^{{\delta_1 k}}}
    \frac{{k}^{{\delta_2 k}}}{\nu^{\nu}{k}^{{\delta_1 k}} n^{\alpha(r\delta k - \nu)}}
    \frac{1}{n^{\frac{\alpha-1}{2}k}}\\
    \leq{} &
    \left(\frac{e^{{\delta_1} + {r\delta} + {\delta_2}} g^{r\delta} \beta^{r\delta} \gamma^{{\delta_2}} {\delta_1}^{{\delta_2}}}{{\delta_1}^{{\delta_1}}}\right)^k
    \sum_{\nu=r\delta k/g}^{r\delta k}
    \frac{{k}^{{\delta_2 k}}}{\nu^{\nu}{k}^{{\delta_1 k}} n^{\alpha(r\delta k - \nu)}}
  \frac{1}{n^{\frac{\alpha-1}{2}k}}\\
  \leq{} &
    \left(\frac{e^{{\delta_1} + {r\delta} + {\delta_2}} g^{2r\delta} \beta^{r\delta} \gamma^{{\delta_2}} {\delta_1}^{{\delta_2}}}{{\delta_1}^{{\delta_1}}}\right)^k
    \sum_{\nu=r\delta k/g}^{r\delta k}
    \frac{{k}^{{\delta_2 k}}}{(r\delta k)^{\nu}{k}^{{\delta_1 k}} k^{\alpha(r\delta k - \nu)}}
  \frac{1}{n^{\frac{\alpha-1}{2}k}}\\
  \leq{} &
    \left(e^{{\delta_1} + {r\delta} + {\delta_2}} g^{2r\delta} \beta^{r\delta} \gamma^{{\delta_2}} {\delta_1}^{{\delta_2}}\right)^k
    \sum_{\nu=r\delta k/g}^{r\delta k}
    \frac{{k}^{{\delta_2 k}}}{k^{\nu}{k}^{{\delta_1 k}} k^{\alpha(r\delta k - \nu)}}
  \frac{1}{n^{\frac{\alpha-1}{2}k}}
\end{align*}
  Let $\psi'$ be the exponent of $k$ in a term of this sum. Then we have
  \begin{align*}
    \psi' &= \delta_2 k - \nu - \delta_1 k - \alpha(r \delta k) + \alpha\nu\\
          &= (1-\alpha)(r \delta k) + (\alpha-1)\nu\\
          &\leq 0.
  \end{align*}
  Using $\phi$ as defined, we arrive at the following inequality:
  \begin{align}
    \sum_{\nu=r\delta k/g}^{r\delta k} \Prob_{\text{dense}}(\delta_1 k,\nu,\delta_2 k)
    \leq \phi^{\frac{(\alpha-1)}{2}k}
    \sum_{\nu=r\delta k/g}^{r\delta k}
    \frac{1}{n^{\frac{\alpha-1}{2}k}}
    \leq{} r\delta k \cdot \left(\frac{\phi}{n}\right)^{\frac{\alpha-1}{2}k}
  \end{align}
  This inequality completes our proof. \qed
\end{proof}

\subsubsection{Density}

Before turning to our main result, we need two more lemmas that establish the probability
of graphs generated using $G(n,m,p)$ having special types of dense subgraphs.

\def\restupper{ e^{g+1} \gamma^g g \beta }
\def\restupperxe{ e^{g+2} \gamma^g g \beta }
\def\rest{ \left(\frac{ \restupper  }{  u/k  }\right)^u }
\begin{theorem}\label{thm:direct-edges}
  Let~$c \geq 1$ be a constant and let~$g := 2\frac{\alpha+c}{\alpha-1}$.
  For $u \leq m, k \leq n$, the probability that the bipartite graph associated
  with $G(n,m,p)$ contains $u$ attributes of degree $\leq g$ that
  generate at least $\rho \geq u$ edges between $k$ fixed vertices is at most
  $$
    \rest \left(\frac{k}{n}\right)^u.
  $$
\end{theorem}

\noindent
We note that it is perhaps surprising that $\rho$ disappears in the upper
bound given above. Since we are assuming that the degree of the attributes is
bounded by $g$, the number of attributes $u$ must be at least $\rho / {g
\choose 2}$. Thus the $\rho$ reappears upon expansion. Since we can bound the
degree of the attributes \whp when $\alpha > 1$, this theorem is generally
applicable to sparse random intersection graphs.

\begin{proof}
  The probability that $u$ attributes of maximal degree $g$ generate at least $\rho \geq u$ edges
  between $k$ fixed vertices can be upper-bounded by
  \begin{equation*}\label{eq:deg-prob}
    {m \choose u} \sum_{d_1,\dots,d_u} \prod_{i=1}^{u} {k \choose d_i} p^{d_i},
  \end{equation*}
  where $d_1,\dots,d_u$ represent all possible choices of the degrees of $u$ attributes
  such that $\sum_{i=1}^u {d_i \choose 2} \geq \rho$ (\ie the
  degrees of the chosen attributes can generate enough edges). Let $D
  = \sum_{i=1}^u d_i$. From Stirling's inequality, it follows that
  \begin{align*}
    {m \choose u} \sum_{d_1,\dots,d_u} \prod_{i=1}^{u} {k \choose d_i} p^{d_i}%
    \leq &~ \frac{(e\beta n^\alpha)^u}{u^u}
           \sum_{d_1,\dots,d_u} \prod_{i=1}^u
           \frac{(e k)^{d_i}}{d_i^{d_i}}
           \left(\frac{\gamma}{n^{(\alpha +1)/2}}\right)^{d_i} \label{eq:change-gamma-2}\\
    =&~ \frac{(e\beta)^u n^{\alpha u}}{u^u}
        \sum_{d_1,\dots,d_u}
        \frac{e^D k^D}{\prod_{i=1}^u d_i^{d_i}}
        \frac{\gamma^D}{n^{\frac{\alpha+1}{2}D}}\\
    \leq &~ \frac{(e\beta)^u (e\gamma)^{gu} n^{\alpha u}}{u^u}
        \sum_{d_1,\dots,d_u}
        \frac{k^D}{n^{\frac{\alpha+1}{2}D}}.
  \end{align*}
  Since each $d_i$ is at most $g$,
  we can upper bound this term by
  \begin{align*}
    {m \choose u} \sum_{d_1,\dots,d_u} \prod_{i=1}^{u} {k \choose d_i} p^{d_i}%
    &\leq \frac{(e^{g+1} \gamma^g \beta)^u n^{\alpha u}}{u^u}
       \cdot \sum_{d_1,\dots,d_u}
       \frac{k^D}{n^{\frac{\alpha +1}{2}D}}\\
       &= \left(\frac{ e^{g+1} \gamma^g \beta  }{  u/k  }\right)^u
       \sum_{d_1,\dots,d_u} \frac{ n^{\alpha u} k^{D-u} }{n^{\frac{\alpha +1}{2}D}}.
  \end{align*}
    We want to show that $(n^{\alpha u} k^{D-u})/(n^{\frac{\alpha
      +1}{2}D})$ is bounded by $(k/n)^{x}$ for some $x \geq u$. Let us first look at the following
  inequality:
  \begin{align*}
    \left(\frac{\alpha +1}{2}\right)D - \alpha u \geq D - u \Leftrightarrow D \geq 2u
  \end{align*}
  Notice that an attribute of degree one generates no edges, thus we
  can assume that all $d_i \geq 2$. It follows that $D \geq 2u$ and
  thus the inequality holds, therefore
  \begin{align*}
    (n^{\alpha u} k^{D-u})/(n^{\frac{\alpha +1}{2}D}) \leq \left(\frac{k}{n}\right)^{D-u}
    \leq \left(\frac{k}{n}\right)^u.
  \end{align*}
  The probability of $u$ attributes generating at
  least $\rho$ edges between $k$ vertices is then at most
  \begin{align*}
       {m \choose u} \sum_{d_1,\dots,d_u} {k \choose d_i} p^{d_i}%
       \leq
       \left(\frac{ e^{g+1} \gamma^g \beta  }{  u/k  }\right)^u
       \sum_{d_1,\dots,d_u} \left(\frac{k}{n}\right)^u.
  \end{align*}
  Finally, since any $d_i$ can be at most $g$, we can get rid of the
  sum by multiplying with a $g^u$ factor:
  \begin{align*}
       \left(\frac{ e^{g+1} \gamma^g \beta  }{  u/k  }\right)^u
       \sum_{d_1,\dots,d_u} \left(\frac{k}{n}\right)^u
       \leq
       \rest
       \left(\frac{k}{n}\right)^u.
  \end{align*} \qed
\end{proof}
The following lemma is a rather straightforward consequence of Theorem~\ref{thm:direct-edges}.
\def\deltaboundA{e^{g+1}\gamma^g g g' \beta}
\begin{lemma}\label{lemma:no-dense-subgraphs}
  Let $c
  \geq 1$ be a constant, $g := 2\frac{\alpha + c}{\alpha - 1}$, $g'
  := {g \choose 2}$ and $\delta > \deltaboundA$. Then the probability
  that $G(n,m,p)$ contains a subgraph of density $\delta$ on
  $k$ vertices is at most
  \[
  \delta k
  \left(\frac{k}{n}\right)^{\frac{\delta k}{g'}}.
  \]
\end{lemma}
\begin{proof}
  By Lemma~\ref{lemma:attr-degree-bound} we can disregard all graphs whose associated
  bipartite graphs have an attribute of degree greater than
  $g$. We can bound the probability with:
  \begin{equation}\label{eq:pdirect}
    \sum_{u=\frac{\delta k}{g'}}^{\delta k} {m \choose u}
    \sum_{d_1,\dots,d_u} {k \choose d_i} p^{d_i},
  \end{equation}
  where $d_1,\dots,d_u$ represent the degrees of the $u$ attributes
  such $\sum_{i=1}^u {d_i \choose 2} \geq \delta k$ (\ie the
  degrees of the $u$ attributes that generate all direct edges).

  Using Theorem~\ref{thm:direct-edges}, the right hand side of Equation \ref{eq:pdirect} is
  bounded by
  \begin{align*}
    \sum_{u=\frac{\delta k}{g'}}^{\delta k}
    \rest \left(\frac{k}{n}\right)^u
    & \leq{}
    \sum_{u=\frac{\delta k}{g'}}^{\delta k}
    \left( \frac{ e^{g+1}\gamma^g g g' \beta }{ \delta } \right)^u
    \left(\frac{k}{n}\right)^u, \label{eq:change-gamma-3}
  \end{align*}
  using the fact that $u/k \geq \delta / g'$.
  Since we set up $\delta \geq \deltaboundA$, we can cancel these terms
  and simplify the above to
  \begin{equation*}\label{eq:pdirect_final}
    \mathbb{P}_{\text{direct}} \leq{}
    \sum_{u=\frac{\delta k}{g'}}^{\delta k}
    \left(\frac{k}{n}\right)^u
    \leq \delta k
    \left(\frac{k}{n}\right)^{\frac{\delta k}{g'}},
  \end{equation*}
  using the fact that $k/n$ is smaller than one. \qed
\end{proof}

\subsubsection{Main Result}

We finally have all the necessary tools to prove the main theorem of
this section.

\begin{theorem}\label{thm:bounded-exp}
  Fix positive constants $\alpha > 1$, $\beta$ and $\gamma$.  Then \whp the
  class of random intersection graphs $G(n,m,p)$ defined by these constants
  has bounded expansion.
\end{theorem}
\begin{proof}
  We show the two conditions of Proposition~\ref{prop:BoundedExpChar}
  are satisfied in Lemma~\ref{lemma:high-degree} and
  Lemma~\ref{lemma:no-dense-minors}, respectively.\qed
\end{proof}

\begin{lemma}\label{lemma:high-degree}
  Let $c \geq 1$ be a constant,
  $g := 2\frac{\alpha + c}{\alpha - 1}$
  $g' := {g \choose 2}$, and
  $\lambda$ be a constant bigger than $\max(2 \restupperxe,~c)$.
  For $G = {\mathcal G}(n,m,p)$, and for all $\epsilon > 0$,
  it holds with probability $O(n^{-c})$~that
  $$
    \frac{1}{|V(G)|} \cdot \left|\left\{v \in V(G)
    \colon \deg(v) \geq \frac{2 \lambda {g'}}{\epsilon}\right\}\right| \leq \epsilon.
  $$
\end{lemma}
\begin{proof}
  By Lemma~\ref{lemma:attr-degree-bound} we can disregard all bipartite graphs
  that have an attribute of degree greater than $g$. Suppose that for some
  $\epsilon$ there exists a vertex set $S$ of size greater than $\epsilon n$
  in which all vertices have degree at least $2\lambda g' /\epsilon$. This assumption
  implies that there exists a set $F$ of edges of size at least $
  \frac{\epsilon n}{2} \frac{2 \lambda g'}{\epsilon} = \lambda g' \cdot n$
  whose members each have at least one endpoint in $S$. Further, since every
  attribute has degree at most $g$ and thus generates at most $g'$ edges,
  there exists a set $F' \subseteq F$ such that
  \begin{enumerate}[label={(\roman{enumi})}]
  \item $|F'| \geq |F|/g' = \lambda n$,
  \item and every $e \in F'$ is generated by at least one attribute
    that generates no other edge in $F'$.
  \end{enumerate}
  The existence of $F'$ follows from a simple greedy procedure: Pick
  any edge from $F$ and a corresponding attribute, then discard at
  most $g'$ edges generated by this attribute. Repeat.

  We now bound the probability that there exists such a set $F'$. Since $F'$
  is generated by exactly $|F'| = \lambda n$ attributes,
  we can apply Theorem~\ref{thm:direct-edges} to obtain the following bound:
  \begin{align*}
    \sum_{k=1}^{n} {n \choose k}
          \left( \frac{\restupper \cdot k}{\lambda n} \right)^{\lambda n}
          \left( \frac{k}{n} \right)^{\lambda n}
    & \leq~
    \sum_{k=1}^{n} \left(\frac{\restupper}{\lambda} \right)^{\lambda n}
                   \frac{n^k e^k}{k^k} \frac{k^{2\lambda n}}{n^{2\lambda n}} \label{eq:change-gamma-1}\\
    & \leq~
    \left(\frac{\restupperxe}{\lambda} \right)^{\lambda n}
    \sum_{k=1}^{n}
    \left( \frac{k}{n} \right)^{2\lambda n-k}
  \end{align*}
  By the choice of $\lambda$, this expression is bounded by
  \begin{align*}
    \frac{1}{2^{\lambda n}} \sum_{k=1}^{n}
    \left( \frac{k}{n} \right)^{2\lambda n-k}
    \leq
    \frac{n}{2^{\lambda n}}
  \end{align*}
  since every element of the sum is smaller than one and the statement follows.
  Note that $n/2^{\lambda n} < 1/n^c$ since $\lambda > c$, \ie this probability converges
  faster than the one proven in Lemma~\ref{lemma:attr-degree-bound}. \qed
\end{proof}

\noindent
We now prove that the second condition of Proposition~\ref{prop:BoundedExpChar}
holds, completing the proof of Theorem~\ref{thm:bounded-exp}.

\def\deltaboundA{e^{g+1}\gamma^g g g' \beta}
\begin{lemma}\label{lemma:no-dense-minors}
  Let $c \geq 1$ be a constant, $g := 2\frac{\alpha + c}{\alpha - 1}$, $g'
  := {g \choose 2}$, $\phi$ be defined as in
  Theorem~\ref{theorem:subdivision-prob} and
  $
    \delta_r > (2r+1) \cdot
    \max\{\deltaboundA,
          ~(c+1)g'\}.
  $
  Then for every $r \in \N^+\!$, for every $0 < \epsilon < e^{-2}$, and
  for every $H \subseteq G = G(n,m,p)$ with $|V(H)| < \epsilon n$, it holds with
  probability $O(n^{-c})$ that $\topgrad_{r}(H) \geq
  \delta_{r}$.
\end{lemma}
\begin{proof}
  By Lemma~\ref{lemma:stable-subdivisions}, if $G$ contains an
  $r$-shallow topological minor of density $\delta_r$, then for some $i
  \in \{0,\dots,2r\}$ there exists a stable $i$-subdivision of density
  $\delta_r/(2r+1)$. We can then bound the probability of a
  $r$-shallow topological minor by bounding the probability of a
  stable $i$-subdivision of density $\delta_r/(2r+1)$.

  From Lemma~\ref{lemma:no-dense-subgraphs} we know that the
  probability of a $0$-shallow topological minor on $k$ nails
  is bounded by
  \begin{align*}
    \binom{n}{k} \delta k \left(\frac{k}{n}\right)^{\frac{\delta k}{g'}}.
  \end{align*}
  By Theorem~\ref{theorem:subdivision-prob}, the density for an
  $i$-subdivision of density $\delta_r/(2r+1)$ for $i \in
  \{1,\dots,2r\}$ is bounded by \[
    r\delta k \cdot \left(\frac{\phi}{n}\right)^{\frac{\alpha-1}{2}k}.
  \]
  Taking the union bound of these two events gives us a
  total bound of
  \begin{align}\label{eqn:unionbound}
    \binom{n}{k} \delta k \left(\frac{k}{n}\right)^{\frac{\delta
        k}{g'}} + (2r+1)r\delta k \cdot
    \left(\frac{\phi}{n}\right)^{\frac{\alpha-1}{2}k}
  \end{align}
  for the probability of a dense subgraph or subdivision on~$k$
  vertices to appear. Taking the union bound over all~$k$ we
  obtain for the first summand that
  \begin{equation*}
  \begin{split}
    \sum_{k=1}^{\epsilon n} \binom{n}{k} \delta k \left(\frac{k}{n}\right)^{\frac{\delta k}{g'}}
    \leq~ \delta_r \sum_{k=1}^{\epsilon n} \frac{n^k e^k}{k^k} \frac{k^{(c+1)k + 1}}{n^{(c+1) k}}.
  \end{split}
  \end{equation*}
  Since $\delta_r$ is a constant, it suffices that the term
  \begin{align*}
    \sum_{k=1}^{\epsilon n} \frac{n^k e^k}{k^k} \frac{k^{(c+1)k + 1}}{n^{(c+1) k}}
  \end{align*}
  is in $O(n^{-c})$. We will show this term is bounded by a geometric sum by considering
  the ratio of two consecutive summands:
  \begin{align*}
    \frac{e^{k+1} (k+1)^{c(k+1) + 1}}{n^{c (k+1)}} \cdot \frac{n^{c
        k}}{e^k k^{ck + 1}}
    ={}& e \frac{(k(1+1/k))^{c(k+1) + 1}}{n^{c} k^{ck + 1}}
    \leq{} e^2 \frac{k^{c}}{n^{c}} \leq e^2 \epsilon^{c}.
  \end{align*}
  Since this summand is smaller than one when $\epsilon < e^{-2}$ and $c \geq
  1$, the summands decrease geometrically; hence its largest element
  (\ie the summand for~$k=1$) dominates the total value of the sum.
  More precisely, there exists a constant $\xi$ (depending on $\alpha$
  and $c$) such that
  \begin{align}\label{eqn:summand1}
    \sum_{k=1}^{\epsilon n} \frac{e^k k^{ck + 1}}{n^{c k}} \leq \xi
    \frac{e}{n^{c}} = O(n^{-c}).
  \end{align}
  We now turn to the second summand. It is easy to see by the same
  methods as before that this sum is also geometric for $n >
  \phi^{(\alpha+1)/2}$ and as such there exists a constant $\xi'$
  which bounds the sum when multiplied with the first element. An
  $r$-shallow topological minor of density $\delta_r$ has at least
  $2\delta_r$ nails, thus we can assume $k \geq 2\delta_r$. Since
  $\delta_r > (c+1)g' \geq c/(\alpha-1)$, we have:
  \begin{align}\label{eqn:summand2}
    \sum_{k={2\delta_r}}^{\epsilon n} (2r+1)r\delta k \cdot
    \left(\frac{\phi}{n}\right)^{\frac{\alpha-1}{2}k} \leq
    \frac{\xi'(2r+1)\phi^{\delta_r}}{n^{(\alpha-1)\delta_r}} \leq
    \frac{\xi'(2r+1)\phi^{\delta_r}}{n^{c}} = O(n^{-c}).
  \end{align}
  Combining (\ref{eqn:summand1}) and (\ref{eqn:summand2}), Equation
  \ref{eqn:unionbound} is bounded by $O(n^{-c})$, as claimed. \qed
\end{proof}


\section{Hyperbolicity}

We now turn to the question of whether the structure of the shortest-path
distances in random intersection graphs is tree-like, using Gromov's
$\delta$-hyperbolicity as defined in Section~\ref{sec:prelim:hyper}. We
establish a negative result by giving a logarithmic lower bound, for all
values of $\alpha$. Our approach is based on a special type of path, which
gives natural lower bounds on the hyperbolicity.

\begin{defn}[$k$-special path]
Let $G = G(n,m,p)$ be a random intersection graph. The $k$-path
$P=v_1,v_2,\ldots,v_{k+1}$ in $G$ is called a $\emph{ $k$-special path}$ if
all the internal vertices of $P$ have degree two in $G$ and there exists
another disjoint path connecting $v_1$ and $v_{k+1}$ in $G$.  We allow for the
second path to have length $0$: this occurs if $P$ is a $k$-cycle such that
all but one vertex of $P$ has degree two in $G$.
\end{defn}

\begin{lemma}\label{lem:k-special}
  Let $k$ be a positive integer and $G = G(n,m,p)$.  If $G$ contains a
  $k$-special path, then $G$ has hyperbolicity at least
  $\lfloor\frac{k}{4}\rfloor$.
\end{lemma}
\begin{proof}
  Let $P = v_1, v_2, \dots, v_{k+1}$ be the $k$-special path in $G$. By definition, $P$ is part of a
  cycle $C$ in $G$; note that $C$ has length at least
  $k$.  We can suppose that the length of $C$ is
  exactly $k$: it will be clear from the remainder of the proof that the lower bound on the hyperbolicity of $G$ increases as the length of $C$ increases.

Setting $v=v_1$ satisfies
  \begin{equation}\label{eq:hyper}
  \forall u\in P_G[v_{\lfloor k/4 \rfloor},v_{\lceil k/2\rceil}]\cup P_G[v_{\lceil k/2\rceil}, v_{\lceil 3k/4\rceil}],\;\; |u-v|_G\geq \lfloor k/4 \rfloor.
  \end{equation}
  Since $v_1\in P[v_{\lfloor k/4 \rceil},v_{\lceil 3k/4 \rceil}]$, (\ref{eq:hyper}) is exactly what is necessary to show that the hyperbolicity of $G$ is at least $\lfloor k/4 \rfloor$.
  \qed
\end{proof}

\noindent
Showing that $k$-special paths exist in an intersection graph is non-trivial,
but crucial for our proof of the following theorem.

\begin{theorem}\label{thm:hyper_main}
  Fix constants $\alpha,\beta$ and $\gamma$ such that $\gamma^2\beta>1$.  There
  exists a constant $\xi>0$ such that \aas, the random intersection graph $G =
  G(n,m,p)$ with $m=\beta n^\alpha$ and $p=\gamma n^{-(1+\alpha/2)}$ has
  hyperbolicity
  \begin{enumerate}[label={(\roman{enumi})}]
    \item at least $\xi \log n$ when $\alpha \geq 1,$
    \item $(1\pm o(1))\xi\log n$ otherwise.
  \end{enumerate}
\end{theorem}

\noindent 
To prove Theorem \ref{thm:hyper_main} we will define another structure
to look for in the bipartite model, which will imply the
existence of $k$-special paths. More specifically, we will restrict
our attention to a particular kind of $k$-special path inside the
giant component of $G$.

\begin{defn}[$k$-special bipartite path]\label{def:k-special-bipartite}
  Let $G = G(n,m,p)$ and $B$ be the associated bipartite graph, fix
  $X\subset V$ and $Y\subset A$.  Letting $B'$ be the subgraph of $B$
  induced by $V\backslash X$ and $A\backslash Y$, we consider a
  connected component $C$ in $B'$. We are interested in paths
  $v_1,v_2,\ldots,v_{2k-1}$ in $B$ such that $v_1,v_{2k-1}$ are both
  elements of $A\backslash Y$ and all the other vertices of the path
  belong to $X\cup Y$.  We will restrict our attention to the paths
  where $v_1$ and $v_{2k-1}$ are both adjacent to vertices of $C$.
  Such a path in $B$ will correspond to a $k$-special path in $G$ if
  the following three conditions hold:

  \begin{enumerate}[label={(\roman{enumi})}]
  \item $N_B(v_1)\cap X=\{v_2\}$ and $N_B(v_{2k-1})\cap
    X=\{v_{2k-2}\}$, \label{condition:endpoint:neighbors:X}
  \item for
    $i=1,2,\ldots,k-2,\;\;N_B(v_{2i+1})=\{v_{2i},v_{2i+2}\}$, \label{condition:attribute:neighbors}
  \item for $i=1,2,\ldots,k-1,\;\;N_B(v_{2i}) \cap
    N_B(V\backslash\{v_{2i}\})=\{v_{2i-1},v_{2i+1}\}$. \label{condition:node:neighbors}
  \end{enumerate}
  We call such paths {\em $k$-special bipartite paths on $(X,Y,C)$}.
\end{defn}

\noindent
Note that when there is no chance of confusion, we may drop $X,Y$ and
$C$ from our notation and merely refer to ``$k$-special bipartite
paths.''

We are now ready to prove Theorem \ref{thm:hyper_main}.  For
convenience, we break up the proof into a lemma for each regime of $\alpha$.\looseness-1

\begin{lemma}\label{lem:hyperbolicity:greater}
  Fix positive constants $\alpha>1$ and $\beta,\gamma$, such that
  $\beta\gamma^2>1$.  Then there exists a constant $\xi>0$ such that
  \aas $G\in G(n,m,p)$ has hyperbolicity at
  least $\xi\log n$.
\end{lemma}

\begin{proof}
  Since $nmp^2>1$ and $\alpha> 1$, we can pick $\zeta>0$ such that
  $(1-\zeta)^2nmp^2>1$.  Let $X\subset V$ be a random subset of size
  $\zeta n$ and $Y\subset A$ of size $\zeta m$.  Consider exposing (or
  inspecting) the edges of $B$ incident with $V\backslash X$ and
  $A\backslash Y$ -- that is, determine exactly which pairs in
  $(V\backslash X)\times(A\backslash Y)$ are edges in $B$.  Suppose
  however, that we do not inspect the edges of $B$ incident with
  either $X$ or $Y$.  We now have a subgraph of $G$ on $V\backslash
  X$.  We call this subgraph the ``{\it exposed graph}.''  Due to our
  choice of $\zeta$, \aas, the exposed graph
  has a giant component of size at least $\delta n$ where the constant
  $\delta =\delta (\alpha,\beta,\gamma,x,n_0)$ for all $n\geq n_0$
  \cite{beh}.  Conditioning on this (likely) event, let $C$ be the
  giant component of the exposed graph.

  Instead of finding (and counting) $k$-special paths in $G$, it will
  be convenient to look for $k$-special bipartite paths on $(X,Y,C)$.
  While each $k$-special bipartite path in $B$ corresponds to a
  $k$-special path in $G$, this correspondence is not one-to-one.
  However, this discrepancy is not a problem since ultimately we will be interested
  in showing that, for an appropriate value of $k$, there is at least
  one $k$-special path in $G$ \aas

  Let $S_k$ denote the number of $k$-special bipartite paths, and
  recall that we are conditioning on the fact that the exposed graph
  has a giant component of size at least $\delta n$. The distribution
  of $S_k$ depends on $n,m,p,\zeta$ and $\delta$.  We approximate the
  first two moments of $S_k$ and then maximize $k$ under the
  constraint that \aas $S_k>0$.  Suppose that
  $v_1$ and $v_{2k-1}$ belong to $A\backslash Y$, with $v_3,\ldots,
  v_{2k-3}\in X$ and $v_2,v_4,\ldots,v_{2k-2}\in Y$, such that
  $v_1$ and $v_{2k-1}$ both have neighbors in $C$.  Denote the
  probability that these vertices form a $k$-special path
  $v_1,v_2,\ldots,v_{2k-1}$ by $p_k$.

  It is convenient to break up the event that the vertices form a
  $k$-special bipartite path into smaller events.  In particular, let
  $\mathbf{P}$ be the event that $B|_{v_1,v_2\ldots,v_{2k-1}}$ is
  exactly a $2k-2$ path on $v_1,v_2,\ldots, v_{2k-1}$.  Let
  $\mathbf{N_1}$ be the event that
  $$N_B(v_1)\cap X=\{v_2\}$$
  and $\mathbf{N_{2k-1}}$ the event that
  $$N_B(v_{2k-1})\cap X=\{v_{2k-2}\}.$$
  Together, these two events correspond to Condition
  \ref{condition:endpoint:neighbors:X} in the definition of $k$-special
  bipartite paths.  For $i=3,5,\ldots,2k-3$, define $\mathbf{N_i}$ to be
  the event
  $$N_B(v_{2i+1})=\{v_{2i},v_{2i+2}\}.$$
  Collectively, these correspond to Condition
  \ref{condition:attribute:neighbors} in the definition of $k$-special
  bipartite paths.  Finally for $i=2,4,\ldots,2k-4$ define $\mathbf{N_i}$ to
  be the event $$N_B(v_{2i}) \cap
  N_B(V\backslash\{v_{2i},v_{2(i+1)},\ldots,v_{2(k-1)}\})=\{v_{2i-1}\}$$ and
  $\mathbf{N_{2k-2}}$ to be the event that $N_B(v_{2i}) \cap
  N_B(V\backslash\{v_{2i}\})=\{v_{2i-1},v_{2i+1}\}$. The event
  $\bigwedge_{i=2,4,\ldots,2k-2}\mathbf{N_i}$ is equivalent to Condition
  \ref{condition:node:neighbors}.  By Lemma \ref{lem:isolates},
  \whp~$$\left|N_B(V\backslash\{v_{2i},v_{2(i+1)},\ldots,v_{2(k-1)}\})\right|\leq
  (1+\epsilon nmp)$$ holds for each $i=1,\ldots,k-1$.  Thus \whp, for
  each $i=1,2,\ldots,k-1$
  \begin{equation}\label{eq:conditional}
    \mathbb{P}\Big[\mathbf{N_{2i}}\Big\vert\;\mathbf{P} \textstyle\bigwedge_{j=1}^{2i-1} \mathbf{N_j} \Big]
    \geq [(1-p)^{(1+\epsilon)nmp}]^{k-1}.
  \end{equation}
  On the other hand, it is clear that $\mathbb{P}[\mathbf{P}]=p^{2k-2}$
  and for $i=1,5,\ldots, k-2$,
  \begin{equation}
    \mathbb{P}\Big[\mathbf{N_{2i+1}}\Big\vert\mathbf{P} \textstyle\bigwedge_{j=1}^{2i} \mathbf{N_j}\Big] = (1-p)^{n-(k-1)}.\label{eq:conditional:2}
  \end{equation}
  Since we know that
  \begin{equation*}
    p_k = \mathbb{P}\Big[\mathbf{P}\textstyle\bigwedge_{i=1}^{2k-1} \mathbf{N_i}\Big] =\mathbb{P}[\mathbf{P}]\prod_{i=1}^{2k-1}\mathbb{P}\Big[\mathbf{N_i}\Big\vert\;\mathbf{P} \bigwedge_{j=1}^{i-1} \mathbf{N_j} \Big],
  \end{equation*}
  we can substitute from Equations \ref{eq:conditional} and
  \ref{eq:conditional:2} to get a lower bound for $p_k$ of
  \begin{equation}\label{eq:pk:substitute}
    p^{2k-2}q^{k^2-3k+2} [q^{|X|-(k-1)}]^2 [q^{n-(k-1)}]^{k-2} [q^{(1+\epsilon)nmp}]^{k-1}
  \end{equation}
  where $q = 1-p$, which simplifies to
  \begin{equation*} \label{eq:pk:last} p_k \geq
    p^{2k-2}q^{(1+\epsilon)(k-1)nmp+(2\zeta+k-2)n-2k+2}.\end{equation*}
  Using the inequality $1-p\geq \exp(-2p)$ (which holds for small enough
  $p$), we have
  \begin{align*}
    p_k & \geq p^{2k-2} \exp\big[-(2p)[(1+\epsilon)(k-1)nmp+(2\zeta +k-2)n-2k+2]\big]\nonumber \\
    & \geq p^{2k-2}
    \exp[-(1+2\epsilon)(k-1)2\beta\gamma^2].
  \end{align*}
  We now count the number of ways, $N_k$, in which a $k$-special path
  could occur in $G$.  By Lemma \ref{lem:isolates}, \whp, the number of
  attributes adjacent to vertices of $C$ is at least $(1-\epsilon)\delta
  nmp$.  Similarly, the number of attributes adjacent to vertices of
  $V\backslash X$ is at most $(1+\epsilon) nmp$.  Thus there are
  $\binom{(1-\epsilon)\delta nmp}{2}$ possible choices for $v_1$ and
  $v_{2k-1}$.  Setting $t = \zeta n$, there are
  $t(t-1)(t-2)\cdots(t-k+2)$ many choices for
  $v_2,v_4,\ldots,v_{2(k-1)}$.  Similarly, setting $s= m-(1+\epsilon)
  nmp$, there are at least $s(s-1)(s-2)\cdots(s-k+3)$ many choices for
  the vertices $v_3,v_5,\ldots, v_{2k-3}$. By linearity of expectation,
  \begin{align*}
    \mathbb{E}[S_k] & \geq p_k \cdot \binom{(1-\epsilon)\delta nmp}{2} [(1-\epsilon)m]^{k-2} (\delta n)^{k-1} (1-o(1)) \nonumber \\
    & = \frac{p_k}{2} \delta^2\beta^2\gamma^2(1-\epsilon)^{2}n^{1+\alpha} [(1-\epsilon)m]^{k-2} (\zeta n)^{k-1} (1-o(1))\nonumber\\
    & = \frac{p_k}{2} \delta^2\beta^2\gamma^2(1-\epsilon)^{k}\zeta^{k-1}n^{(k-1)(1+\alpha)+1}(1-o(1)) \nonumber\\
    & =  \exp\big[-(1+2\epsilon)(k-1)2\beta\gamma^2\big] \delta^2\beta^2\gamma^2(1-\epsilon)^{k}\zeta^{k-1}n(1-o(1)) \nonumber\\
    & =  n\delta^2\beta^2\gamma^2 \exp\left[k\log(1-\epsilon)+ \big( -2(1+2\epsilon)\beta\gamma^2 +\log \zeta\big)(k-1)\right](1-o(1)) \nonumber \\
    & =
    \frac{n(\delta\beta\gamma e^{\beta\gamma^2(1+2\epsilon)})^2}{\zeta}\exp\left[k
      \big(\log(1-\epsilon) -2(1+2\epsilon)\beta\gamma^2 +\log
      \zeta\big)\right](1-o(1)).
  \end{align*}
  Thus there exists a positive constant $\xi$ such that
  $\mathbb{E}[S_{\xi \log n}]=\omega(1)$, namely:
  $$ \xi < \frac{-1}{\log(1-\epsilon)-2(1+2\epsilon)\beta\gamma^2+\log \zeta}.$$
  Note that the denominator is also negative, so we can indeed pick
  $\xi>0$.

  We now show that $S_k$ is tightly concentrated around its mean for
  the values of $k$ when $\mathbb{E}[S_k]=\omega(1)$. Denote $S_k$ as
  the sum of $n_k$ random indicator variables
  $\mathcal{I}_{v_1,v_2,\ldots,v_{2k-1}}$ where
  $\mathcal{I}_{v_1,v_2,\ldots,v_{2k-1}}=1$ if there is a $k$-special
  bipartite path on the vertices $v_1,v_2,\ldots,v_{2k-1}$.  We would
  like to calculate $\mathbb{P}[\mathcal{I}_{u_1,u_2,\ldots,u_{2k-1}}=1
  | \mathcal{I}_{v_1,v_2,\ldots,v_{2k-1}}]$.  If
  $\{v_1,v_2,\ldots,v_{2k-1}\}$ and $\{u_1,u_2,\ldots,u_{2k-1}\}$ are
  not disjoint then the probability is $0$.  Otherwise
  $\mathbb{P}[\mathcal{I}_{u_1,u_2,\ldots,u_{2k-1}}=1 |
  \mathcal{I}_{v_1,v_2,\ldots,v_{2k-1}}]\leq p_k/(1-p)^{2k(k-1)}$, since the
  event $\mathcal{I}_{v_1,v_2,\ldots,v_{2k-1}}$ implies that
  $$N_B(\{v_2,v_4,\ldots,v_{2(k-1)}\})\cap N_B(\{u_2,u_4,\ldots,u_{2(k-1)}\})=\emptyset.$$
  We conclude that
  \begin{align}
    Var(S_k)&\leq n_kp_k +n_k(n_k-1)\frac{p_{k,v}^2}{(1-p)^{2k(k-1)}} - (n_kp_k)^2 \nonumber\\
    &\leq n_k p_k +(n_k p_k)^2\bigg[\frac{1}{(1-p)^{2k(k-1)}}-1\bigg] \nonumber\\
    &\leq n_k p_k +(n_k p_k)^2\big[e^{-4pk(k-1)}-1\big] \label{eq:var:greater}\\
    &\leq n_k p_k +(n_k
    p_k)^2[8pk(k-1)]. \label{eq:var:lesser}
  \end{align}
  Inequality \ref{eq:var:greater} follows from the fact that
  $1-e^x\geq e^{-2x}$ for $0<x<1$, while Inequality \ref{eq:var:lesser}
  follows from the fact that $e^x\leq 1+2x$ for $0<x<\log 2$.  Thus, for
  $t>0$ and $k=O(\log n)$, by Chebyshev's inequality,
  \begin{align}
    \mathbb{P}[|S_k-\mathbb{E}[S_k]|\geq t\mathbb{E}[S_k]]&\leq \frac{Var(S_k)}{t^2\mathbb{E}[S_k]^2}\nonumber\\
    &= \frac{n_k p_k}{(tn_k p_k)^2} +8pk(k-1)t^{-2} \nonumber\\
    &\leq \frac{1}{t^2n_kp_k}+8pk(k-1)t^{-2}=o(1)\nonumber.
  \end{align}
  Therefore we have shown \aas that
  $S_k=\mathbb{E}[X](1\pm o(1))$ when $k=O(\log n)$. In particular we
  have shown that there exists a positive constant $\xi$ such that
  \aas $G$ has hyperbolicity at least $\xi \log
  n$.
\qed\end{proof}

\begin{lemma}\label{lem:hyperbolicity:equal}
  Fix positive constants $\alpha,\beta,\gamma$, such that $\alpha=1$ and
  $\beta\gamma^2>1$.  Then there exists a constant $\xi>0$ such that
  \aas $G\in G(n,m,p)$ has hyperbolicity at least
  $\xi\log n$.
\end{lemma}
\begin{proof}
  The proof is very similar to the proof for the case $\alpha>1$.
  Pick $\zeta>0$ such that $(1-\zeta)^2nmp^2>1$.  Now let $X$ be a
  subset of $\zeta n$ vertices and $Y$ a subset of $\zeta m$
  attributes.  Consider the subgraph $B'$ of $B$ induced on
  $V\backslash X\times A\backslash Y$.  Let $G'$ be the subgraph of
  $G$ derived from the bipartite graph $B'$; $G'$ will be the
  ``exposed graph.''  Note that since $(1-x)^2nmp^2>1$, \aas~the exposed graph has a giant component $C$ of size
  $\delta n$ for an appropriate constant $c$~\cite{Lageras_A-note_2008}.

  We again restrict our inquiry to the existence of $k$-special
  bipartite paths on $(X,Y,C)$.  Let $S_k$ be the number of
  $k$-special bipartite paths (conditioning on the fact that the
  exposed graph has a giant component of size at least $\delta n$).
  Given the vertices $v_1,v_2,\ldots,v_{2k-1}$ such that $v_1$ and
  $v_{2k-1}$ both have neighbors in $C$, and such that
  $v_3,v_5,\ldots,v_{2k-3}\in Y$ and $v_{2},v_4,\ldots,v_{2k-2}\in Y$,
  we would like to know the probability, denoted $p_k$, that the
  vertices form a $k$-special bipartite path $P$:
  $v_1,v_2,\ldots,v_{2k-1}$.  In this setting, Equation
  \ref{eq:pk:substitute} becomes
  \begin{align*}
    p_k &\geq
    p^{2k-2}(1-p)^{k^2-3k+2}[(1-p)^{|X|-k+1}]^2[(1-p)^{n-k+1}]^{k-2}[(1-p)^{m-k}]^{k-1}\\
        &\geq p^{2k-2} [(1-p)^{(k-1)(m-k)+(k-2)n+2\zeta n-2k+2}] \nonumber \\
    &\geq p^{2k-2} [(1-p)^{(\beta+1)nk+(2\zeta-2-\beta)n -k^2-k+2}] \nonumber \\
    &\geq p^{2k-2} \exp(-2p\big[(\beta+1)nk+\gamma(2\zeta-2-\beta)n -k^2-k+2\big])\\
    &\geq p^{2k-2} \exp(-2\big[\gamma k(\beta+1)+
      \gamma(2\zeta-2-\beta)\big])(1-o(1)),
  \end{align*}
  where the last inequality holds for $k=o(n)$.

  Again, we count the ways in which a $k$-special bipartite path can
  occur.  Setting $t = \zeta n$, there are $t(t-1)(t-2)\cdots(t-k+2)$--many
  choices for $v_2,v_4,\ldots,v_{2(k-1)}$.  Similarly, setting
  $s= \zeta m$, there are $s(s-1)(s-2)\cdots(s-k+3)$--many choices for
  the vertices $v_3,v_5,\ldots, v_{2k-3}$. By Lemma \ref{lemma:neighbors},
  \whp~there are at least $(1-\epsilon)\beta\delta n$ attributes
  adjacent to vertices of $C$.  Thus, by linearity of expectation,

\begin{align*}
  \mathbb{E}[X_k] &\geq \frac{1}{2}p_k(\zeta n)^{k-1}(\zeta m)^{k-2}((1-\epsilon)\beta\delta n)^{2} (1-o(1))  \\
  \begin{split}
    &\geq\frac{1}{2}n (\gamma^2\zeta^2\beta)^{k-1} ((1-\epsilon)\beta\delta )^{2} \\
    & \qquad\cdot\exp(-2[ k(\gamma\beta+\gamma)+ \gamma(2\zeta-2-\beta)])(1-o(1))
  \end{split}\\
  \begin{split}
    &\geq \frac{((1-\epsilon)\beta\delta
      )^{2}}{(\gamma^2\zeta^2\beta)}n \\
    & \qquad\cdot\exp(-2[ k(\gamma\beta+\gamma
    +\log(\gamma^2\zeta^2\beta))+ \gamma(2\zeta-2-\beta) ])(1-o(1)).
  \end{split}
\end{align*}
Thus there is a $\xi>0$ such that $\mathbb{E}[S_{\xi\log n
  -1}]=\omega(1)$; namely, any $\xi$ satisfying
$$\xi< \frac{1}{2\beta\gamma+\gamma+2\log(\gamma^2\zeta^2\beta)}.$$
The proof that $S_k$ is tightly concentrated around its mean is
exactly the same as in the case when $\alpha>1$, so we omit it here.
Therefore we have finished the proof for $\alpha = 1$.
\qed\end{proof}

\begin{lemma}\label{lem:hyperbolicity:less}
  Fix positive constants $\alpha,\beta,\gamma$, such that $\alpha<1$
  and $\beta\gamma^2>1$.  Then there exists a constant $\xi>0$ such
  that \aas $G\in G(n,m,p)$ has hyperbolicity
  at least $\xi\log n$.
\end{lemma}

\begin{proof}
  In essence, the proof of Lemma \ref{lem:hyperbolicity:greater}
  consists in showing that the bipartite graph $B$, associated with
  $G$ \aas, contains a $2\lfloor k/4
  \rfloor$-slim triangle, i.e.\ that there exist three vertices $x$,
  $y$, and $z$ with shortest paths $P[x,y],P[y,z]$, and $P[x,z]$
  between them such that
$$\exists v\in P[x,y]: \forall w\in P[x,z]\cup P[z,y], d_G(v,w)\leq 2\lfloor k/4\rfloor.$$
This fact is then used to show that $G$ has a $\lfloor k/4\rfloor$-slim
triangle.

Note that the bipartite graphs which define the intersection graphs
$G(n,m,p)$ and $G(m,n,p)$ have the same distribution: In the first
case the intersection graph is formed by projecting onto the vertices of
the bipartite graph, while in the second case the projection is onto
the attributes.  It is not hard to see that if the bipartite graph $B$
contains a $2\delta$-slim triangle then both of the two possible
projections will contain $\delta$-slim triangles.  Thus the proof of
Lemma \ref{lem:hyperbolicity:less} follows directly from the proof of
Lemma \ref{lem:hyperbolicity:greater}.
\qed\end{proof}

\renewcommand{\topfraction}{0.8}
\renewcommand{\textfraction}{0.25} 
\section{Experimental evaluation}\label{sec:experiments}

Our theoretical results provide insight into asymptotic properties of the
grad, degeneracy, and hyperbolicity of random intersection graphs. To sharpen
our understanding of how these statistics behave in realistic parameter
ranges, we designed four experiments to relate our theoretical predictions to
concrete measurements.

We used the NetworkX Python package~\cite{networkx} to generate our random
intersection graphs (using the \texttt{uniform\_random\_intersection\_graph}
method), and the SageMath software system~\cite{sage} to compute the
hyperbolicity~\cite{sageHypA,sageHypB,sageHypC},
degeneracy~\cite{sageDegA,sageDegB} and
diameter~\cite{sageDiamA,sageDiamB,sageDiamC,sageDiamD} of the generated
graphs. The measurements of the $p$-centered coloring number (presented below)
were executed using the implementation available in~\cite{concuss}. In the
first three experiments, we generated random intersection graphs using
parameters~$\alpha \in \{0.3,0.5,0.7,0.9,1.0,1.2\}$ and fixed~$\beta = \gamma
= 1.2$.  Each data point represents an average over~20 random instances of a
given size $n$ (increasing from a few thousand to several hundred thousand,
with finer granularity at smaller sizes to capture boundary effects). The last
experiment, which concerns the structural sparseness of~$G(n,m,p)$ in the
regime~$\alpha > 1$, fixes parameters~$\alpha = 1.5$, $\beta = 0.1$
and~$\gamma = 5$ (due to computational constraints), and averages over ten
instances of each size.

Our first experiment is designed to estimate the constants involved in the
asymptotic bounds provided by Theorem~\ref{thm:degen_main}. To that end, we
fit the three functions for the respective regimes of~$\alpha$ by computing a
multiplicative scaling $\tau$ using least-square fitting via the
\texttt{scipy}~\cite{scipy} implementation of the Levenberg--Marquardt
algorithm~\cite{levenberg,marquardt}.
\begin{figure}[p]
	\centering
	\includegraphics[width=\linewidth]{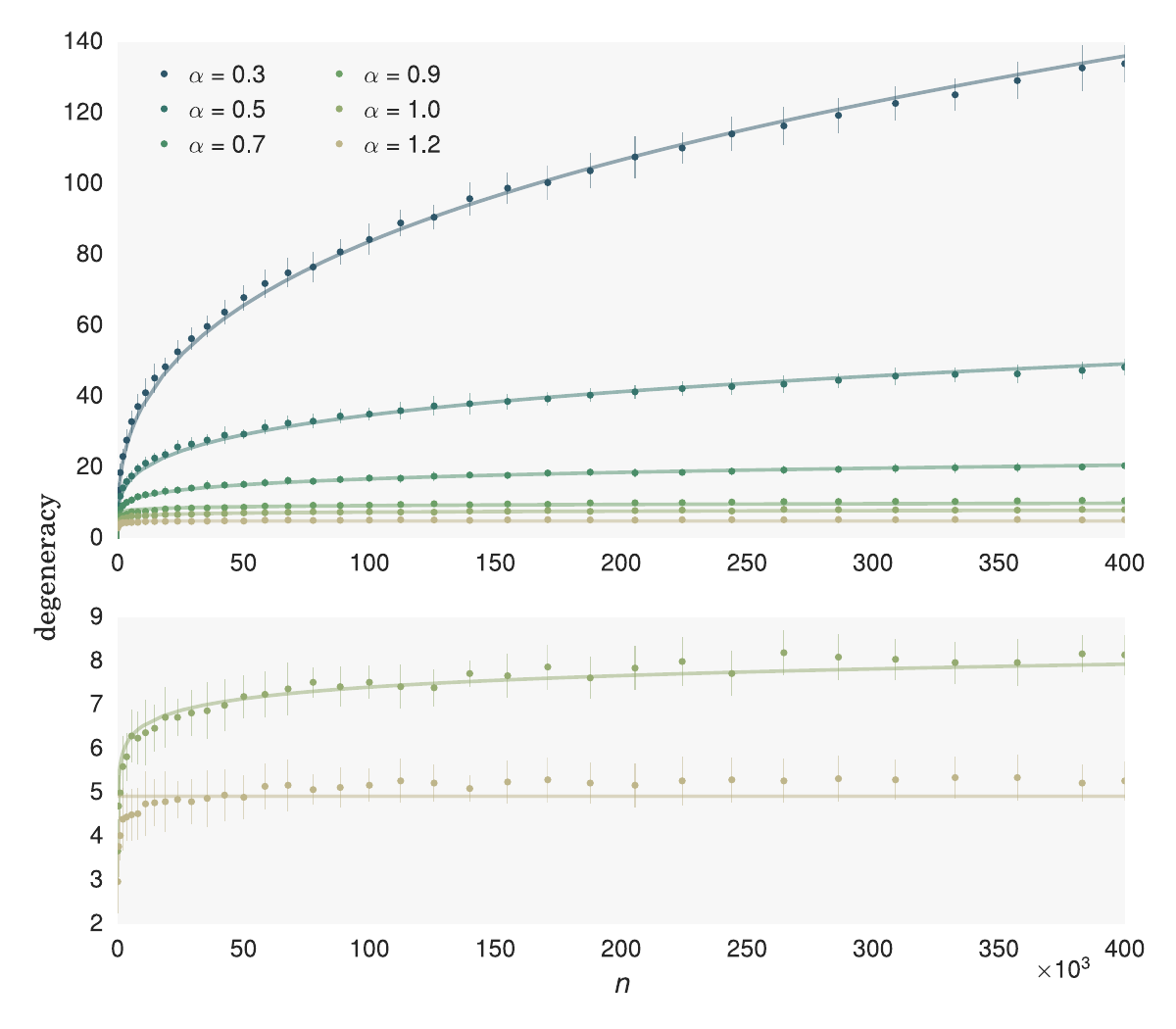}
	\caption{\label{fig:degeneracy}%
      Degeneracy of~$G(n,m,p)$ for different values of~$\alpha$ and
      increasing~$n$.  The parameters~$\beta = \gamma = 1.2$ were fixed; all data points are averaged
      over 20 graphs. Error bars show one standard deviation.
      The lower figure contains the same plots for~$\alpha \geq 1$ in a different scale.
      The continuous lines are functions listed in Table~\ref{table:fitted}
      fitted to the data.}
\end{figure}
\begin{table}[p]
    \hspace*{15pt}
    \begin{minipage}[c]{0.33\textwidth}\vspace{0pt}
        \begin{tabular}{l>{\hspace*{1em}} l>{\hspace*{1em}} l}
            \toprule
            \multicolumn{1}{c}{$\alpha$} &
                \multicolumn{1}{c}{Function} &
                    \multicolumn{1}{c}{$\tau$} \\ \midrule
            0.3 & $\tau \cdot 1.2 n^{0.35}$ & 1.24 \\
            0.5 & $\tau \cdot 1.2 n^{0.25}$ & 1.63 \\
            0.7 & $\tau \cdot 1.2 n^{0.15}$ & 2.49 \\
            0.9 & $\tau \cdot 1.2 n^{0.05}$ & 4.34 \\
            1.0 & $\tau \cdot \frac{\log n}{\loglog n}$ & 1.57 \\
            1.2 & $\tau$ & 4.92 \\ \bottomrule
        \end{tabular}
    \end{minipage}%
    \begin{minipage}[c]{0.62\textwidth}
        \caption{\label{table:fitted}%
            Functions corresponding to the degeneracy upper-
            and lower bounds from Theorem~\ref{thm:degen_main}
            fitted to the degeneracy data displayed in Figure~\ref{fig:degeneracy}.
            The coefficients~$\tau$ were determined by least-square fitting.
        }
    \end{minipage}
    \hfill
\end{table}
Both the data and the fitted functions are plotted in
Figure~\ref{fig:degeneracy}, the function parameters and scaling factors can
be found in Table~\ref{table:fitted}. Already for graphs of moderate size, we
see that the degeneracy closely follows the predicted functions. We further note
that for the series~$\alpha = 1.2$, the observed degeneracy is around 5, which is
very far from the massive upper bound given by
setting $r = 0$ in Lemma~\ref{lemma:no-dense-minors} (value not shown in plot).
It would be interesting to see whether bounds with tighter constants can be
obtained by different proof techniques. For the value~$\alpha = 1.0$, we see
that the asymptotic lower bound~$\Omega(\frac{\log n}{\loglog n})$ fits the
observed degeneracy very well with only a small scaling factor of~$1.57$. We
put forward the conjecture that the degeneracy actually
follows~$\Theta(\frac{\log n}{\loglog n})$ in this regime. Finally,
for~$\alpha < 1$ we see some increase of the scaling factor~$\tau$ as~$\alpha$
tends to one. The lower bound~$\gamma n^{(1-\alpha)/2}$ therefore seems to
miss some slight dependency on~$\alpha$, but otherwise matches the degeneracy
observed very well.

\begin{figure}[t]
    \centering
    \includegraphics[width=\linewidth]{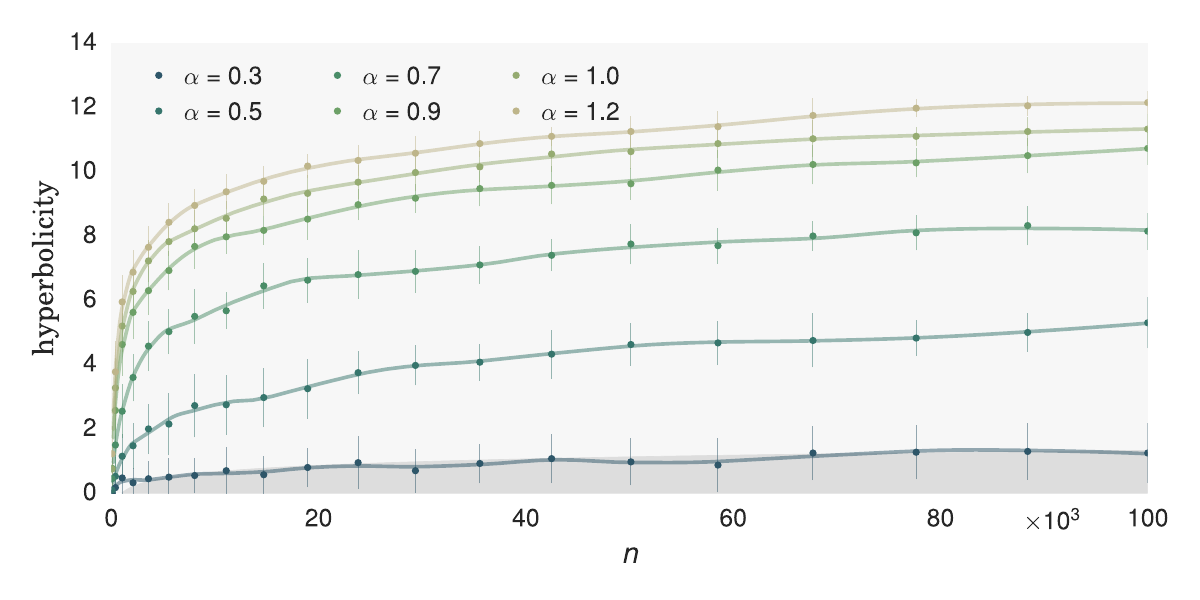}
    \caption{\label{fig:hyperbolicity}%
      Hyperbolicity and of~$G(n,m,p)$ for different
      values of~$\alpha$ and increasing~$n$. The
      parameters~$\beta = \gamma = 1.2$ were fixed; each point represents an average
      over~20 graphs. Error bars show one standard deviation.
      The darker grey area marks the theoretical lower bound~$0.298 \log n$.
      Lines represent smoothed versions of the series and are included as a visual guide.}
\end{figure}

We designed our second experiment to see how tight our lower bound of ~$\xi \log
n$ is for the hyperbolicity of~$G(n,m,p)$. The value
of~$\xi$ for~$\gamma = \beta = 1.2$ turns out to be close to~$0.194$ for~$\alpha = 1$,
and~$0.289$ for $\alpha \neq 1$. Figure~\ref{fig:hyperbolicity} contains the
results for the same~$\alpha$, $\beta$, $\gamma$ values as before for graphs
up to size~$10^5$. We can see quite clearly that the lower bound~$0.298 \log
n$ is rather pessimistic for larger values of $\alpha$. Only at~$\alpha = 0.3$ do
we see a plot that follows this lower bound tightly. This observation suggests that a more
fine-grained analysis could provide not only tighter lower bounds but likely
very tight matching upper bounds as well.

\begin{figure}[t]
    \centering
    \includegraphics[width=\linewidth]{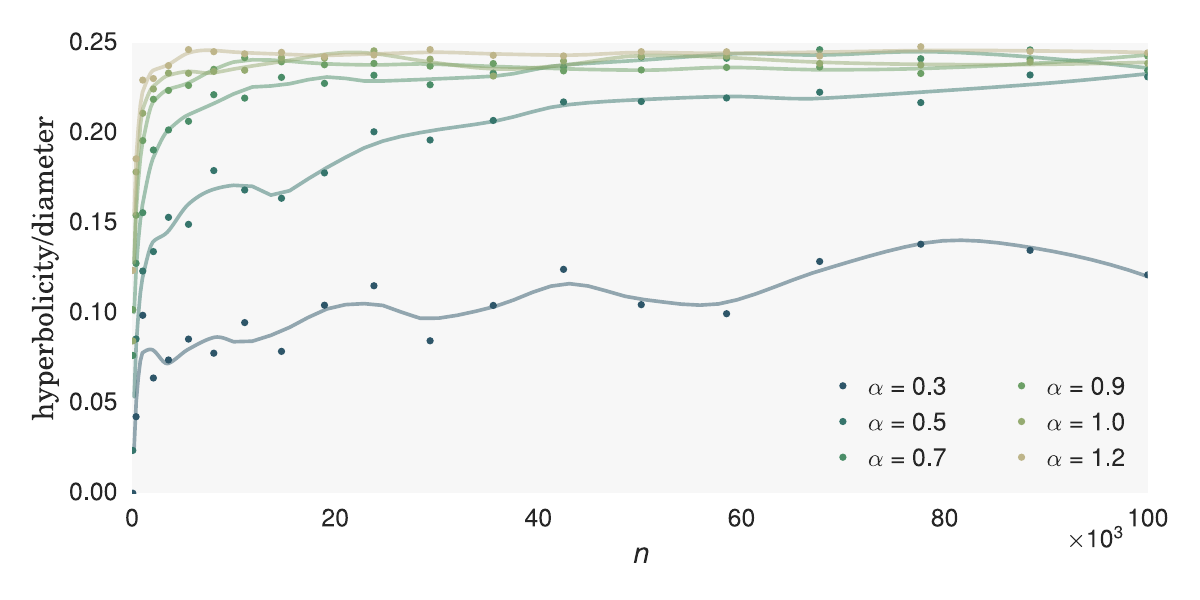}
    \vspace*{-3em}
    \caption{\label{fig:hyp-vs-diam}%
      Ratio of hyperbolicity and diameter of~$G(n,m,p)$ for different
      values of~$\alpha$ and increasing~$n$. The
      parameters~$\beta = \gamma = 1.2$ were kept constant and all
      data points were averaged over~20 graphs. Due to error propagation, the
      error bars are large and we omit them for the sake of clarity. Lines represent
      a smoothed version of the series and are included as a visual guide.}
\end{figure}

The third experiment is related to the second: here we tested the relationship
between the diameter and the hyperbolicity of~$G(n,m,p)$.
Figure~\ref{fig:hyp-vs-diam} plots the ratio\footnote{For disconnected graphs, we use the values
from the largest connected component.} of hyperbolicity to diameter, and
appears to show convergence to constants depending on~$\alpha$. As expected,
hyperbolicity and diameter seem to be asymptotically related by a constant
factor. However, the plots also reveal a periodic fluctuation for smaller
graphs that disappears as the graph size increases.

Finally, our last experiment measures the structural sparseness
of~$G(n,m,p)$ in the regime~$\alpha > 1$. Since our bounds on the
degeneracy---the most `local' grad~$\topgrad_0$---are far away from what we
observed in the first experiment, it is reasonably to presume that the
bounds on higher grads are even worse. Since bounded expansion has large
potential to be exploited algorithmically in practice, we want to obtain a
better understanding of the orders of magnitudes involved.

\begin{table}[t]
    \small
    \begin{minipage}[c]{0.35\linewidth}\vspace{0pt}
        \caption{\label{table:asymptote}%
            Estimated asymptotic values of the plots in
            Figure~\ref{fig:hyp-vs-diam}, obtained by averaging
            the series' last~$11$ values. As seen on the right, the
            values fit the logistic function
            $.24 (1+2^{-15.21(\alpha-.31)})^{-1}$ well (residuals plotted in lower portion).
        }
    \end{minipage}\hspace*{.5em}%
    \begin{minipage}[c]{0.18\linewidth}\vspace{0pt}
        \centering
        \begin{tabular}{l>{\hspace*{1em}} l}
            \toprule
            \multicolumn{1}{c}{$\alpha$} &
                \multicolumn{1}{c}{$\hat d$}  \\ \midrule
            0.3 & 0.115 \\
            0.5 & 0.213 \\
            0.7 & 0.237 \\
            0.9 & 0.237 \\
            1.0 & 0.240 \\
            1.2 & 0.245 \\ \bottomrule
        \end{tabular}
    \end{minipage}%
    \begin{minipage}[c]{0.45\linewidth}
        \includegraphics[width=\linewidth]{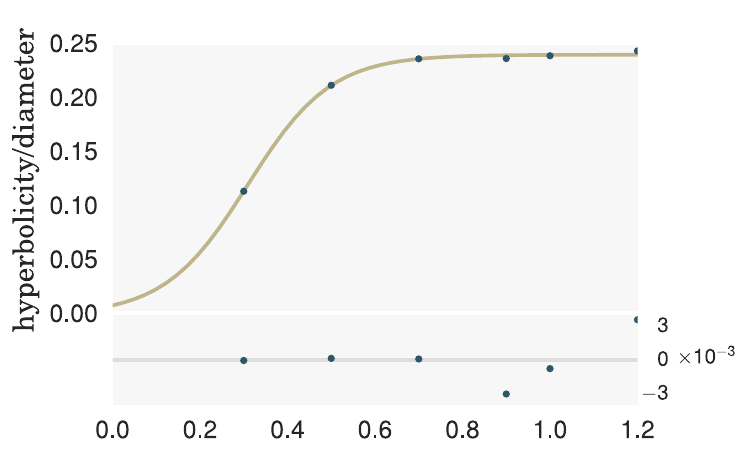}
    \end{minipage}
    \vrule
\end{table}

The asymptotic bounds provided by Lemma~\ref{lemma:no-dense-minors} are
incredibly pessimistic: For parameters $\alpha=1.5$, $\gamma=5$
and $\beta=0.1$ (selected to be relatively realistic and enable
easy generation) the bound on~$\topgrad_r$ provided by this lemma
is at least $10^{13}$ (independent of $r$) even if we only insist on
an error probability of~$O(n^{-1})$.
Since all tools for classes of bounded expansion depend
heavily on the behavior of the expansion function and the expansion function
given by the framework in~\cite{NOdM08} will depend on $\delta_r$, this upper
bound is not enough to show practical applicability. Our experiment
provides empirical evidence that the upper
bound is not tight, improving the prospects for these associated tools.
Specifically, we calculate so-called \emph{$p$-centered colorings}, which can
be used to characterize classes of bounded expansion and have immediate algorithmic
applications~\cite{NOdM08}.

\begin{proposition}[$p$-centered colorings~\cite{NOdM08}]\label{theorem:low-td-coloring}
  A graph class $\mc G$ has bounded expansion if and only if there
  exists a function $f$ such that for every $G \in \mc G$, $p \in \N$,
  the graph $G$ can be colored with $f(p)$ colors so that any $i < p$
  color classes induce a graph of treewidth $\leq i$ in $G$.  This
  coloring can be computed in linear time.\looseness-1
\end{proposition}

\begin{figure}[h]
  \includegraphics[width=\textwidth]{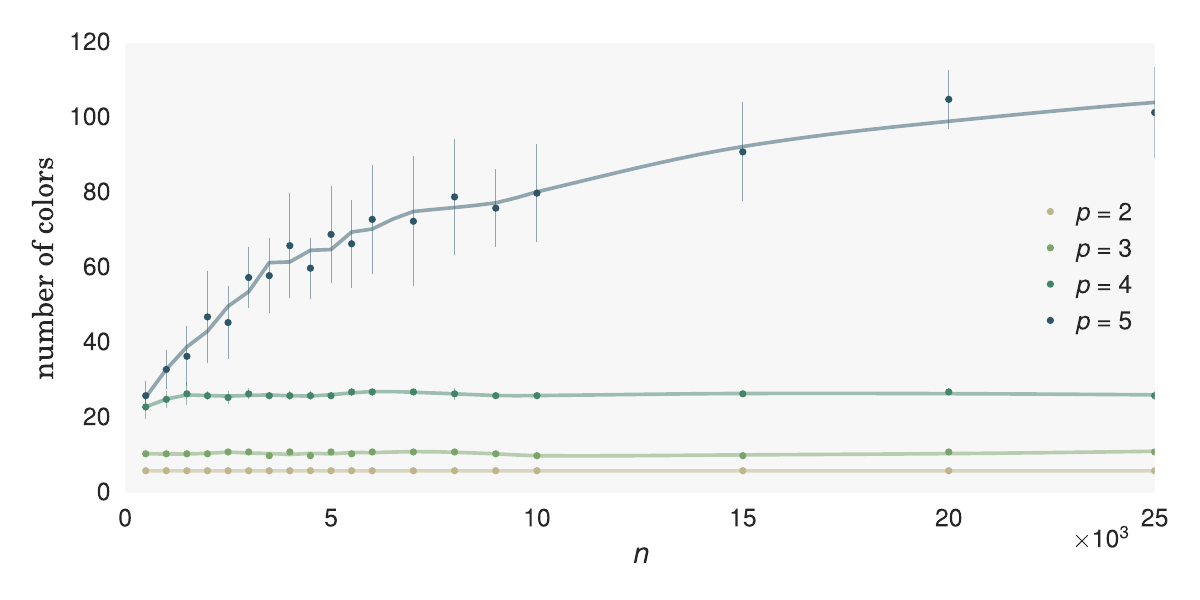}
  \caption{\label{fig:expdata}%
    Median number of colors in a $p$-centered coloring for~$G(n,m,p)$
    with parameters $\alpha = 1.5$, $\beta = 0.1$ and $\gamma
    = 5$ (taken over ten random instances). Error bars denote one standard deviation
    (for $p\leq 4$ hardly visible). Lines represent a smoothed versions of the series and are included as a visual guide.}
\end{figure}

\noindent
The characterization can be made stronger (the colorings are actually
low \emph{treedepth} colorings), but this fact is not important in this context.
We implemented a simple version of the linear time coloring algorithm
and ran it on ten random intersection graphs for each
($n \in \{500$, $1000$, $\dots$, $6000$, $7000$,$\dots$,
$10,000$, $15,000$, $20,000$, $25,000\}$) with parameters
$\alpha=1.5$, $\gamma=5$ and $\beta=0.1$ for each $p \in \{2, 3, 4, 5\}$.
Figure~\ref{fig:expdata} shows the median number of colors used by the algorithm.
Our theoretical results predict a horizontal asymptote
for every $p$. We can see a surprisingly small bound for $p \in
\{2,3,4\}$. Even for $p=5$ the plot starts flattening within the experimental
range. It should be noted that the colorings given by this simple
approximation algorithm are very likely to be far from optimal (\ie the colorings
may use many unnecessary colors).\looseness-1

This result indicates that the graphs modeled by random intersection are
amenable to algorithms based on $p$-centered colorings (which usually perform
dynamic programming computations that depend exponentially on the number of
colors). Further, by the known relation between $p$-centered colorings and
the expansion function, these experiments indicate that these graphs have much more reasonable
expansion bounds than Lemma~\ref{lemma:no-dense-minors} would suggest.

\section{Conclusion and open problems}

In this paper we have determined the conditions under which random
intersection graphs exhibit two types of algorithmically useful structure.
We proved graphs in $G(n,m,p)$ are structurally sparse (have bounded expansion)
precisely when the number of attributes in the associated bipartite graph
grows faster than the number of vertices ($\alpha > 1$). Moreover, we showed that
when the generated graphs are not structurally sparse, they fail to achieve
even much weaker notions of sparsity (in fact, \whp they contain large cliques).

On the other hand, we showed that the metric structure of random intersection
graphs is not tree-like for any value of $\alpha$: the hyperbolicity (and
treelength) grows at least logarithmically in $n$.  While we only determine a
lower bound for the hyperbolicity, we believe this bound to be the correct order of
magnitude since the diameter (a natural upper bound for the hyperbolicity) of a
similar model of random intersection graphs was shown to be $O(\log
n)$~\cite{Rybarczyk_Diameter_2011}. Our experimental results support this
hypothesis: the ratio of hyperbolicity to diameter seems to converge to a
constant.

A question that naturally arises from these results is if structural sparsity
should be an expected characteristic of practically relevant random graph
models. Our contribution solidifies this idea and supports previous results
for different random graph models~\cite{BndExpNetw14,FelixThesis}.
We further ask whether the grad is small enough to enable practical
algorithmic application---our empirical evaluation using $p$-centered colorings
of random intersection graphs with $\alpha > 1$ indicate the
answer is affirmative.

\bigskip
\small{\noindent
\textbf{Acknowledgments:} The authors would like to thank Kevin
Jasik of RWTH Aachen University for generating the data for the $p$-centered coloring experiment.
Portions of this research are a
product of work started during the ICERM research cluster ``Towards
Efficient Algorithms Exploiting Graph Structure'', co-organized by
B. Sullivan, E. Demaine, and D. Marx in April 2014.
N. Lemons funded by the Department of Energy at Los Alamos National Laboratory
under contract DE-AC52-06NA25396 through the Laboratory-Directed Research
and Development Program.
F. S\'anchez Villaamil funded by DFG-Project RO 927/13-1 ``Pragmatic Parameterized Algorithms''.
B.~D.~Sullivan supported in part by DARPA GRAPHS/SPAWAR Grant
N66001-14-1-4063,  the Gordon \& Betty Moore Foundation under DDD Investigator
Award GBMF4560, and the National Consortium for Data Science.
Any opinions, findings, and conclusions or recommendations expressed in this
publication are those of the author(s) and do not necessarily reflect the
views of DARPA, SSC Pacific, DOE, the Moore Foundation, or the NCDS.
}
\bibliography{biblio}

\end{document}